\documentclass[preprint2]{aastex}
\usepackage{color}
\usepackage{epsfig}
\usepackage[fleqn]{amsmath}
\usepackage{multirow}
\usepackage{mathptmx}
\usepackage{graphicx}
\usepackage{bm}

\usepackage{lineno}

\shorttitle{Construction of explicit symplectic integrators}
\shortauthors{Wang et al.}

\begin{document}


\title{Construction of explicit symplectic integrators in general relativity.
III. Reissner-Nordstr\"{o}m-(anti)-de Sitter black holes}


\author{Ying Wang$^{1,2}$, Wei Sun$^{1}$, Fuyao Liu$^{1}$, Xin Wu$^{1,2,3,\dag}$}
\affil{1. School of Mathematics, Physics and Statistics, Shanghai
University of Engineering Science, Shanghai 201620, China
\\ 2. Center of Application and Research of Computational Physics,
Shanghai University of Engineering Science, Shanghai 201620, China
\\  3. Guangxi Key Laboratory for Relativistic Astrophysics, Guangxi
University, Nanning 530004, China} \email{Emails:
wangying424524@163.com (Y. W.), sunweiay@163.com (W. S.),
liufuyao2017@163.com (F. L.); $\dag$ Corresponding Author:
wuxin$\_$1134@sina.com (X. W.)}


\begin{abstract}

We give a possible splitting  method to a Hamiltonian for the
description of charged particles moving around the
Reissner-Nordstr\"{o}m-(anti)-de Sitter black hole with an
external magnetic field. This Hamiltonian can be separated into
six analytical solvable pieces, whose solutions are explicit
functions of proper time. In this case, second- and fourth-order
explicit symplectic integrators are easily available. They exhibit
excellent long-term behavior in maintaining the boundness of
Hamiltonian errors regardless of ordered or chaotic orbits if
appropriate step-sizes are chosen. Under some circumstances,  an
increase of positive cosmological constant gives rise to
strengthening the extent of chaos from  the global phase space;
namely, chaos of charged particles occurs easily for the
accelerated expansion of the universe. However, an increase of the
magnitude of negative cosmological constant does not. The
different contributions on chaos are because the cosmological
constant acts as a repulsive force in the
Reissner-Nordstr\"{o}m-de Sitter black hole, but an attractive
force in the Reissner-Nordstr\"{o}m-anti-de Sitter black hole.

\end{abstract}


\emph{Unified Astronomy Thesaurus concepts}: Black hole physics
(159); Computational methods (1965); Computational astronomy
(293); Celestial mechanics (211)




\section{Introduction}
\label{sec:intro}

Many exact solutions on black holes can be derived from Einstein's
general relativistic field equations. The first statical
spherically symmetric solution is the Schwarzschild metric
(Schwarzschild 1916). When the mass source in the origin has an
electric charge, the Schwarzschild metric is the famous
Reissner-Nordstr\"{o}m (RN) solution (Reissner 1916; Nordstr\"{o}m
1918). When a positive constant $\Lambda$ for describing the
static cosmological model introduced by Einstein is included in
the Schwarzschild metric, the Schwarzschild-de Sitter metric is
provided. With the inclusion of cosmological constant $\Lambda$,
the Reissner-Nordstr\"{o}m-(anti)-de Sitter [RN-(A)dS] solution
(Reissner 1916; Kottler 1918; Nordstr\"{o}m 1918) is also
obtained.

A small positive cosmological constant acts as a gravitationally
repulsive energy component, and models dark energy for dominating
the energy budget of the universe leads to an accelerated
expansion of the universe (Ashtekar 2017). This result is strongly
supported by observational evidences (Riess et al. 1998;
Perlmutter et al. 1999; Ade et al. 2016). In fact, the
cosmological constant describes a weak field gravity for the
dynamics of galaxies, galaxy groups and clusters, and is supported
by the parameters of galactic halos and of a sample of galaxy
groups (Gurzadyan 2019; Gurzadyan $\&$ Stepanian 2019).
Surprisingly, the theory of gravitational waves from isolated
systems links up with the presence of a positive cosmological
constant (Ashtekar et al. 2016; Ashtekar 2017).  Although the
known observational results do not  support  the existence of a
negative cosmological constant, the negative cosmological constant
becomes useful in the study of the black hole thermodynamics under
the existence of quantum fields in gravitational fields.
Appropriate boundary conditions  should be imposed in black holes
so that the black holes remain thermally stable. The AdS boundary,
as a reflecting wall, can cause asymptotically AdS black holes to
remain thermally stable. The phase transition between
Schwarzschild AdS black holes and thermal AdS space was discovered
by Hawking $\&$ Page (1983). Finite temperature configurations in
the decoupled field theory are associated to black hole
configurations in AdS spacetimes. The black holes yield the
Hawking radiation in the AdS spacetime (Maldacena 1998). On the
other hand,  the cosmological constant affects some properties of
accretion discs orbiting black holes in quasars and active
galactic nuclei (Stuchl\'{i}k 2005). These accretion discs are
developed through the accretion process of dust and gases around
compact objects. The mass of the accreting body increases during
the accretion process, but it could also decrease (Jamil et al.
2008; Jamil 2009). Karkowski $\&$ Malec (2013) found that the
damping effect on the mass accretion rate is weaker for positive
values of the cosmological constant than for negative values of
the cosmological constant, and it is related to the ratio of the
dark energy density to the fluid density. The authors mainly
focused on the question of how the cosmological constant exerts
influences on the possible accretion rate. Ficek (2015) considered
relativistic Bondi-type accretion in the RN-(A)dS spacetime.

The Schwarzschild, RN and RN-(A)dS spacetimes are highly
nonlinear, but they are integrable and non-chaotic. Magnetic
fields can destroy the integrability of these spacetimes. In fact,
a strong magnetic field with hundred Gauss is indeed present in
the vicinity of the supermassive black hole at the Galactic center
(Eatough et al. 2013). The presence of an external magnetic field
was shown through high-frequency quasiperiodic oscillations and
relativistic jets observed in microquasars (Stuchl\'{i}k et al.
2013; Kolo\v{s} et al. 2017).  It can influence oscillations in
the accretion disk and the creation of jets. The accretion of
charged particles from the accretion disk goes towards the black
hole due to a radiation-reaction force. The external magnetic
field that is usually considered  is so weak that it does not
perturb a spacetime metric; namely, the external magnetic field
does not appear in the spacetime metric and is not a solution of
the field equations of general relativity. However, such weak
magnetic field will essentially affect motions of electrons or
protons when the ratios of charges of electrons or protons to
masses of electrons or protons are large (Tursunov et al. 2018).
Of course, some magnetic fields can perturb spacetime metrics,
such as the magnetized Ernst-Schwarzschild spacetime geometry
(Ernst 1976). The chaotic motions of charged particles around the
Schwarzschild and Kerr black holes were confirmed in a series of
papers (Nakamura $\&$ Ishizuka 1993; Takahashi $\&$ Koyama 2009;
Kop\'{a}\v{c}ek et al. 2010; Kop\'{a}\v{c}ek $\&$ Karas 2014;
Stuchl\'{i}k $\&$ Kolo\v{s} 2016; Kop\'{a}\v{c}ek $\&$ Karas 2018;
P\'{a}nis et al. 2019; Stuchl\'{i}k et al. 2020; Yi $\&$ Wu 2020).
In particular, the chaotic dynamics of neutral particles in the
magnetized Ernst-Schwarzschild spacetim is possible due to the
magnetic field acting as a gravitational effect although no
electromagnetic forces exist (Karas $\&$ Vokroulflick\'{y} 1992;
Li $\&$ Wu 2019). Besides external magnetic fields, spin of
particles is an important source for inducing chaos. Chaos can
occur when a spinning test particle moves in the spacetime
background of a Kerr black hole (Lukes-Gerakopoulos et al. 2016).
Resonances and chaos for spinning test particles in the
Schwarzschild and Kerr backgrounds were shown by Brink et al.
(2015a, b) and Zelenka et al. (2020). Some  spacetime backgrounds,
e.g.,  the Manko, Sanabria-G\'{o}mez, Manko (MSM) metric can be
chaotic (Dubeibe et al. 2007; Han 2008; Seyrich $\&$
Lukes-Gerakopoulos 2012). The word ``chaos" is a nonlinear
behavior of a dynamical system and describes the system with
exponentially sensitive dependence on initial conditions. The
chaotic or regular feature of orbital motions can be reflected on
the gravitational waves (Zelenka et al. 2019).

Detecting the chaotical behavior requires that a numerical
integration scheme have good stability and precision, i.e.,
provide reliable results to the trajectories. The Schwarzschild,
RN and RN-(A)dS spacetimes are integrable, but their solutions are
unknown. Thus,  numerical schemes are still necessary to solve
these systems. It is well known that symplectic integrators are
the most appropriate solvers for long-term evolution of
Hamiltonian systems. Clearly, Hamiltonians of the Schwarzschild,
RN and RN-(A)dS spacetimes are not separations of variables.
Perhaps each of the Hamiltonians may be split into two integrable
parts, but no explicit functions of proper time can be given to
analytical solutions of the two parts in the general case.
Explicit symplectic integrators (Swope et al. 1982; Wisdom 1982;
Ruth 1983; Forest $\&$ Ruth 1990; Wisdom $\&$ Holman
1991\footnote{The Wisdom-Holman method is not strictly explicit
because an implicit iterative method is required to calculate the
eccentric anomaly in the solution of the Kepler main part.}) that
have been developed in the solar system and share good long-term
numerical performance in the preservation of geometric structures
and constants of motion\footnote{The preservation of constants of
motion (such as energy integral) in symplectic integrators does
not mean that the constants  are exactly conserved, but
 means that errors of the constants are bounded. The
energy-preserving integrators (Bacchini et al. 2018a, b; Hu et al.
2019, 2021) and the projection methods (Fukushima 2003a,b,c, 2004;
Ma et al. 2008; Wang et al. 2016, 2018; Deng et al. 2020) are not
symplectic, whereas they do exactly conserve energy.} become
useless in this case. However, completely implicit or explicit and
implicit combined symplectic integrators should be available
without doubt (Brown 2006; Preto $\&$ Saha 2009; Kop\'{a}\v{c}ek
et al. 2010; Lubich et al. 2010; Zhong et al. 2010; Mei et al.
2013a, 2013b; Tsang et al. 2015). In most cases, explicit
algorithms have advantage over  same order implicit schemes  at
the expense of computational time. Considering this point, we
proposed second- and fourth-order explicit symplectic integration
algorithms for the Schwarzschild spacetime by separating the
Hamiltonian of this spacetime into four parts with analytical
solutions as explicit functions of proper time in Paper I (Wang et
al. 2021a). When five parts similar to the four separable pieces
in the Hamiltonian of the Schwarzschild metric are given to a
splitting form of the Hamiltonian of the RN black hole with an
external magnetic field, explicit symplectic integrators were also
set up in Paper II (Wang et al. 2021b). In other words, a fact
concluded from Papers I and II is that the construction of
explicit symplectic integrations is possible for a curved
spacetime when a Hamiltonian of the spacetime is separated into
more parts with analytical solutions as explicit functions of
proper time. How to split the Hamiltonian depends on what this
spacetime is.

Following Papers I and II, we  want to know in this paper how
explicit symplectic integrators are designed for the RN-(A)dS
spacetime with an external magnetic field. We are particularly
interested in investigating how the cosmological constant effects
the regular and chaotic dynamical behavior of charged particles
moving around the black hole. For the sake of these purposes, we
introduce the relativistic gravitational field in Sect. 2. In
Sect. 3, we give new explicit symplectic integrators and check
their numerical performance. Then, a new explicit symplectic
method with an appropriate step size is applied to survey the
effect of the cosmological constant on the dynamics of orbits of
charged particles. Finally, the main results are concluded in
Sect. 4.

\section{RN black hole with cosmological constant and external
magnetic field}

In dimensionless spherical-like  coordinates $(t, r, \theta,
\phi)$, the RN black hole with electric charge $Q$ and
cosmological constant $\Lambda$ describing the expansion of the
universe is given in the following spacetime (Ahmed et al. 2016)
\begin{eqnarray}
-\tau^2 &=& ds^{2} = g_{\alpha\beta}dx^{\alpha}dx^{\beta}
\nonumber \\
&=& -(1-\frac{2}{r}+\frac{Q^2}{r^2} -\frac{\Lambda}{3}r^{2})dt^{2} \nonumber \\
& & +(1-\frac{2}{r}+\frac{Q^2}{r^2} -\frac{\Lambda}{3}r^{2})^{-1}
dr^{2} \nonumber \\
& &  +r^{2}d \theta^{2} +r^{2}\sin^{2} \theta d \phi^2.
\end{eqnarray}
Here, geometrized units $c=G=M=1$ are given to the speed of light
$c$, the constant of gravity $G$, and the mass of black hole $M$.
Dimensionless operations act on proper time $\tau$, coordinate
time $t$, separation $r$, charge $Q$, and cosmological constant
$\Lambda$. Namely, $\tau\rightarrow M\tau$, $t\rightarrow Mt$,
$r\rightarrow Mr$, $Q\rightarrow MQ$ and
$\Lambda\rightarrow\Lambda/M^2$. The four-velocity
$\mathbf{U}=U^{\mu}=\dot{x}^{\mu}$ is also dimensionless and
satisfies the constraint
\begin{equation}
\mathbf{U}\bullet\mathbf{U}=U^{\alpha}U_{\alpha}=g_{\mu\nu}\dot{x}^{\mu}\dot{x}^{\nu}=-1.
\end{equation}
Unlike the Euclid inner product, the Riemann inner product
$\bullet$ in Eq. (2) can be negative.

$\Lambda=0$ indicates that Equation (1) is an asymptotically flat
metric. A non-vanishing cosmological constant $\Lambda$ represents
a weak field gravity proper for the dynamics of galaxies, galaxy
groups and clusters. $\Lambda>0$ and $\Lambda<0$ correspond to
asymptotically de Sitter (dS) and anti-de Sitter (AdS) spacetimes,
respectively. Without loss of generality, we consider a relation
between realistic and dimensionless values of positive
cosmological constant. The  realistic value of cosmological
constant obtained by Planck data is $\Lambda_{PL}=1.11\times
10^{-52}m^{-2}=1.11\times 10^{-46}km^{-2}$ (Ade et al. 2016).
Various  realistic values of cosmological constant for dark matter
considering the data on galaxy systems from galaxy pairs to galaxy
clusters were given in the range $\Lambda=10^{-52}m^{-2}\sim
10^{-50}m^{-2}$ (Gurzadyan $\&$ Stepanian 2019). Suppose that the
value $\Lambda_{0}=10^{-46}km^{-2}$  considered in a paper of Xu
$\&$ Wang (2017) is regarded as a  realistic value of cosmological
constant, a dimensionless value of cosmological constant can be
estimated in terms of $\Lambda=\Lambda_{0}M^2G^2/c^4$. Obviously,
it depends on the mass of a black hole. For the supermassive black
hole candidate in the center of the giant elliptical galaxy M87
having mass  $M=(6.5\pm 0.7)\times 10^{9}M_{\odot}$ with
$M_{\odot}$ being the Sun's mass (EHT Collaboration et al. 2019a,
b, c), we obtain dimensionless cosmological constant
$\Lambda=\Lambda_{0}M^2G^2/c^4=10^{-46}km^{-2}\times [(6.5\pm
0.7)\times 10^{9}M_{\odot}G/c^2]^2=10^{-46}km^{-2}\times(6.5\pm
0.7)^2\times 10^{18}\times 1.475^2km^2=7.3\times10^{-27}
\sim1.1\times10^{-26}$. Such a small dimensionless value does not
change the spacetime geometry of black hole if the separation $r$
is not large enough. The effect of cosmological constant on the
spacetime should also be negligible for the $150M_{\odot}$ binary
black hole merger GW190521 (Abbott et al. 2020a, 2020b). The
dimensionless value of cosmological constant increases with an
increase of the black hole's mass. If the black hole's mass is
$M=(1/1.475)\times 10^{20}M_{\odot}$, we have dimensionless
cosmological constant $\Lambda=10^{-6}$. For $M=(1/1.475)\times
10^{21}M_{\odot}$, $\Lambda=10^{-4}$. Given
$M=2.14391705774127\times 10^{22}M_{\odot}$, $\Lambda=0.1$. When
$M=(1/1.475)\times 10^{23}M_{\odot}$, $\Lambda=1$. These larger
dimensionless values should exert explicit influences on the
spacetime geometries. The larger dimensionless values of
cosmological constant (e.g. $\Lambda=-0.075, -0.1, 0.1$) in the
references (Ahmed et al. 2016; Shahzad $\&$ Jawad 2019; Shafiq et
al. 2020) may be taken for the existence of extremely supermassive
black holes. In what follows, various values of cosmological
constant are dimensionless. Equation (1) is called the RN-(A)dS
spacetime. It has at most three event horizons (Ahmed et al. 2016;
Shahzad $\&$ Jawad 2019).

Seen from  the expression $g_{00}=-1-2\varphi$ ($\varphi$ is the
gravitational potential) in Eq. (1), the potential $-1/r$ plays a
role in an attractive force proportional to $-1/r^2$. Similarly,
the cosmological constant makes the potential $-\Lambda r^{2}/6$
yield an attractive force proportional to $\Lambda r/3$ for
$\Lambda<0$, but a repulsive force for $\Lambda>0$. The force is
independent of the black hole's mass. However, the black hole's
charge leads to the potential $Q^2/(2r^{2})$ giving a repulsive
force proportional to $Q^2/r^{3}$. As the distance increases, the
attraction from the heavy body and the repulsive force from the
charge decrease; the latter is smaller than the former. However,
the attraction or repulsiveness from the cosmological constant
grows. It eventually dominates the other two forces when the
distance is large enough.

The metric corresponds to a Lagrangian system
\begin{equation}
\mathcal{L} = \frac{1}{2} (\frac{ds}{d\tau})^2
=\frac{1}{2}g_{\mu\nu}\dot{x}^{\mu}\dot{x}^{\nu}.
\end{equation}
It defines a covariant generalized momentum
\begin{equation}
p_{\mu} = \frac{\partial \mathcal{L}}{\partial
\dot{x}^{\mu}}=g_{\mu\nu}\dot{x}^{\nu}.
\end{equation}
Since the Lagrangian does not explicitly depend on  $t$ and
$\phi$, the Euler-Lagrangian equations show that two constant
momentum components with respect to proper time $\tau$ are
\begin{eqnarray}
p_{t} &=& -(1-\frac{2}{r}+\frac{Q^2}{r^2}-\frac{\Lambda}{3}r^{2})\dot{t}=-\mathbb{E},\\
p_{\phi} &=& r^{2}\sin^{2}\theta\dot{\phi}=\mathbb{L}.
\end{eqnarray}
$\mathbb{E}$ is the energy of a test particle moving in the
gravitational field, and $\mathbb{L}$ denotes the angular momentum
of the particle.

Transform the Lagrangian into a Hamiltonian
\begin{eqnarray}
H &=& \mathbf{U}\bullet\mathbf{p}-\mathcal{L}
=\frac{1}{2}g^{\mu\nu}p_{\mu}p_{\nu} \nonumber \\
&=& -\frac{1}{2}(1-\frac{2}{r}+\frac{Q^2}{r^2}
-\frac{\Lambda}{3}r^{2})^{-1} \mathbb{E}^{2} \nonumber \\
&&  +\frac{1}{2}(1-\frac{2}{r}
+\frac{Q^2}{r^2}-\frac{\Lambda}{3}r^{2})p^{2}_{r}
+\frac{1}{2}\frac{p^{2}_{\theta}}{r^2} \nonumber \\
&& +\frac{1}{2}\frac{\mathbb{L}^{2}}{r^2\sin^2\theta}.
\end{eqnarray}
Due to Eq. (2), this Hamiltonian is always identical to a given
constant
\begin{equation}
H=-\frac{1}{2}.
\end{equation}
In fact, a second integral (Carter 1968) excluding $E$ and
$\mathbb{L}$ can be found by the variables separated in the
Hamilton-Jacobi equation. Thus, this system is integrable.

Let the black hole suffer from the interaction of an external
magnetic field. The electromagnetic field is described by a
four-vector potential with two nonzero covariant components
(Felice $\&$ Sorge 2003; Wang $\&$ Zhang 2020)
\begin{eqnarray}
A_t &=& -\frac{Q}{r}, \\
 A_{\phi} &=& \frac{B}{2}g_{\phi\phi}=\frac{B}{2}r^{2}\sin^{2}
\theta.
\end{eqnarray}
$A_t$ is the Coulomb part of the electromagnetic potential. It
does not appear in the metric $g_{\alpha\beta}$, but it should be
given because a charged particle and the black hole's charge $Q$
yields a Coulomb force. This Coulomb force is compensated by
$A_t$. $B$ denotes the strength of the magnetic field parallel to
the $z$ axis. As claimed in the Introduction, $B$ is so weak that
it does not affect the metric but affects the motion of charged
particles. The motion of particles with charge $q$ around the
magnetized black hole is determined by the Hamiltonian
\begin{eqnarray}
K &=& \frac{1}{2}g^{\mu\nu}(p_{\mu}-qA_{\mu})(p_{\nu} -qA_{\nu})
\nonumber \\
&=& -\frac{1}{2}(1-\frac{2}{r}+\frac{Q^2}{r^2}
-\frac{\Lambda}{3}r^{2})^{-1}
(E-\frac{\tilde{Q}}{r})^2 \nonumber \\
&& +\frac{1}{2}(1-\frac{2}{r} +\frac{Q^2}{r^2}
-\frac{\Lambda}{3}r^{2})p^{2}_{r}
+\frac{1}{2}\frac{p^{2}_{\theta}}{r^2} \nonumber \\
&& +\frac{L^{2}}{2r^2\sin^2\theta}+\frac{1}{8}\beta^2
r^{2}\sin^{2} \theta-\frac{\beta L}{2},
\end{eqnarray}
where $Q^{*}=qQ$ and $\beta=qB$. Dimensionless operations (see
also Yi $\&$ Wu 2020) are given as follows: $K\rightarrow Km^2_p$,
$E\rightarrow Em_p$, $p_r\rightarrow p_rm_p$,
$p_{\theta}\rightarrow p_{\theta}Mm_p$, $L\rightarrow LMm_p$,
$B\rightarrow B/M$ and $q\rightarrow qm_p$, where $m_p$ is the
particle's mass.  Unlike $\mathbb{E}$ and $\mathbb{L}$ in Eqs. (5)
and (6), the energy $E$ and angular momentum $L$ are written as
\begin{eqnarray}
E &=& (1-\frac{2}{r}+\frac{Q^2}{r^2}-\frac{\Lambda}{3}r^{2})
\dot{t}+\frac{Q^{*}}{r},
\\
 L &=& r^{2}\sin^{2}\theta\dot{\phi}+\frac{1}{2}\beta
r^{2}\sin^{2} \theta.
\end{eqnarray}
The magnetic field causes the radial potential $\beta^2 r^{2}/8$
to yield  an attractive force proportional to $-\beta^2 r/4$. The
attraction grows with the distance. However, the radial potential
$L^{2}/(2 r^{2})$ from the angular momentum gives a repulsive
force proportional to $L^{2}/r^{3}$. In fact, this repulsive force
is a centrifugal force of inertia. As Wang et al. (2021b) claimed,
the Coulomb term gives the particle a repulsive force effect for
$Q^{*}> 0$, but a gravitational force effect for $Q^{*}<0$. It is
more important than the repulsive force effect from the black
hole's charge. Although the potential from the black hole's charge
and the potential from the angular momentum yield the repulsive
forces, the latter repulsive forces vary their directions when the
charged particles' motions are not restricted to the radial
motions. Similarly, the attractive forces from the magnetic field
unlike those from the negative cosmological constant (or those
from the heavy body) have also different directions in the general
case.

The Hamiltonian $K$ is still equal to the given constant
\begin{equation}
K=-\frac{1}{2}.
\end{equation}
However, no second integral excluding the two integrals $E$ and
$L$ exists in the case. Thus, this Hamiltonian is non-integrable.
Its dynamics is complicated and mainly employs numerical
techniques to be investigated.

\section{Numerical investigations}

First, new explicit symplectic integrators are specifically
designed for the Hamiltonian system (11). Then, the performance of
the new algorithms is evaluated. Finally, the dynamics of order
and chaos of orbits in the Hamiltonian is surveyed, and the
relation between the chaotic behavior and the cosmological
constant is mainly discussed.

\subsection{Construction of explicit symplectic integrators}

In Papers I and II (Wang et al. 2021a, b), the explicit symplectic
integrators were proposed for the  Hamiltonian of Schwarzschild's
metric with four analytical integrable splitting parts,  and the
Hamiltonian of RN's metric with five analytical integrable
separable parts. These successful examples in general relativity
tell us that the key of the construction of an explicit symplectic
integrator lies in splitting the considered Hamiltonian in an
appropriate way. Not only the splitting parts in this Hamiltonian
should be analytically solvable, but also all analytical solutions
in the separable parts should be explicit functions of proper time
$\tau$. Noticing this idea, we begin our work.

Split the Hamiltonian (11) into  six parts
\begin{equation}
K=K_1+K_2+K_3+K_4+K_5+K_6.
\end{equation}
The sub-Hamiltonians are expressed as
\begin{eqnarray}
K_1 &=& -\frac{1}{2}(1-\frac{2}{r}+\frac{Q^2}{r^2}
-\frac{\Lambda}{3}r^{2})^{-1} (E-\frac{Q^{*}}{r})^{2} \nonumber \\
&& +\frac{L^{2}}{2r^2\sin^2\theta}+\frac{1}{8}\beta^2
r^{2}\sin^{2} \theta-\frac{\beta L}{2}, \\
K_{2} &=& \frac{1}{2}p^{2}_{r},\\
K_{3} &=& -\frac{1}{r}p^{2}_{r},\\
K_{4} &=& \frac{p^{2}_{\theta}}{2r^2}, \\
K_{5} &=& \frac{1}{2}\frac{Q^2}{r^2}p^{2}_{r}, \\
K_{6} &=& -\frac{\Lambda}{6}r^2p^{2}_{r}.
\end{eqnarray}
$K_2$, $K_3$, $K_4$ and  $K_5$ are consistent with those in the
decomposing method of the RN black hole in Paper II (Wang et al.
2021b).

$K_1$, $\cdots$, $K_6$ correspond to differential operators
$\mathcal{A}$, $\mathcal{B}$, $\mathcal{C}$, $\mathcal{D}$,
$\mathcal{E}$ and $\mathcal{F}$, respectively. These operators are
written in the forms
\begin{eqnarray}
\mathcal{A} &=& -\frac{\partial K_1}{\partial r}\frac{\partial
}{\partial p_r}-\frac{\partial K_1}{\partial \theta}\frac{\partial
}{\partial p_{\theta}} \nonumber \\
&=&
[-(E-\frac{Q^{*}}{r})^{2}(\frac{1}{r^2}-\frac{Q^{2}}{r^3}-\frac{\Lambda}{3}r)
\nonumber \\ && \cdot
(1-\frac{2}{r}+\frac{Q^2}{r^2}-\frac{\Lambda}{3}r^2)^{-2}
\nonumber \\
&&+\frac{Q^{*}}{r^2}(E-\frac{Q^{*}}{r})
(1-\frac{2}{r}+\frac{Q^2}{r^2}-\frac{\Lambda}{3}r^2)^{-1}
\nonumber \\
&&+\frac{L^{2}}{r^3\sin^2\theta} -\frac{\beta^2}{4}r\sin^2\theta
]\frac{\partial
}{\partial p_r} \nonumber \\
&& + (\frac{L^2\cos\theta}
{r^{2}\sin^{3}\theta}-\frac{1}{8}\beta^2
r^{2}\sin(2\theta))\frac{\partial
}{\partial p_{\theta}} \nonumber  \\
&=& f_1(r,\theta)\frac{\partial }{\partial
p_r}+f_2(r,\theta)\frac{\partial
}{\partial p_{\theta}}, \\
\mathcal{B} &=& p_{r}\frac{\partial}{\partial r}, \\
\mathcal{C} &=& -\frac{2}{r}p_r\frac{\partial}{\partial r}
-\frac{p^{2}_r}{r^2}\frac{\partial}{\partial p_r}, \\
\mathcal{D} &=& \frac{p_{\theta}}{r^{2}}\frac{\partial}{\partial
\theta}+\frac{p^2_{\theta}}{r^{3}}\frac{\partial}{\partial p_r}, \\
\mathcal{E} &=&  \frac{Q^2}{r^2}p_r\frac{\partial}{\partial
r}+Q^2\frac{p^{2}_r}{r^3}\frac{\partial}{\partial p_r}, \\
\mathcal{F} &=& - \frac{\Lambda}{3}r^2p_r\frac{\partial}{\partial
r}+\frac{\Lambda}{3}rp^2_r\frac{\partial}{\partial p_r}.
\end{eqnarray}
Using them to act on the solutions $\textbf{z}(0)=(r_0, \theta_0,
p_{r0},  p_{\theta0})$ at the beginning of proper time step $h$,
we can obtain their corresponding analytical solutions
$\textbf{z}=(r, \theta, p_{r},  p_{\theta})$ over a  step size
$h$. All the analytical solutions are indeed explicit functions of
the proper time step $h$. For example, the analytical solutions
$\textbf{z}$ for the sub-Hamiltonian $K_1$ are expressed in an
exponential operator $e^{h\mathcal{A}}$ as $\textbf{z}=(p_{r},
p_{\theta}) = e^{h\mathcal{A}}\textbf{z}(0)$. That is,
$p_r=p_{r0}+h f_1(r_0,\theta_0)$ and $p_{\theta}=p_{\theta0}+h
f_2(r_0,\theta_0)$. These exponential operators symmetrically
compose  a second-order explicit symplectic integrator
\begin{eqnarray}
S^{K}_2(h) &=& e^{\frac{h}{2}\mathcal{F}}\circ
e^{\frac{h}{2}\mathcal{E}}\circ e^{\frac{h}{2}\mathcal{D}}\circ
e^{\frac{h}{2}\mathcal{C}}\circ e^{\frac{h}{2}\mathcal{B}}\circ
e^{h\mathcal{A}} \nonumber \\
&&\circ e^{\frac{h}{2}\mathcal{B}}\circ
e^{\frac{h}{2}\mathcal{C}}\circ e^{\frac{h}{2}\mathcal{D}}\circ
e^{\frac{h}{2}\mathcal{E}}\circ e^{\frac{h}{2}\mathcal{F}}.
\end{eqnarray}
This explicit symplectic method is specifically designed for the
Hamiltonian (11). It is easily used to yield a fourth-order
symplectic scheme of Yoshida (1990)
\begin{equation}
S^{K}_4(h)=S^{K}_2(\gamma h)\circ S^{K}_2(\delta h)\circ
S^{K}_2(\gamma h),
\end{equation}
where $\delta=1-2\gamma$ and $\gamma=1/(2-\sqrt[3]{2})$.

Here are three points illustrated. (i) the Hamiltonian (11)
becomes the RN Hamiltonian (7) for the case of $\beta=0$. It is
natural that the newly established algorithms are appropriate for
the RN problem. (ii) The splitting technique to the Hamiltonian
(11) may not be unique. Perhaps there may be other splitting
methods that satisfy the requirement for the construction of such
similar explicit symplectic integration algorithms. (iii) The
present operator splitting method can be applied to an extended
version of the Hamiltonian (11) as follows:
\begin{eqnarray}
\Gamma &=&
f_1(r,\theta)+g^{rr}p^2_r+g^{\theta\theta}p^2_{\theta\theta},  \\
g^{rr}&=& (a_0+a_1r+\cdots+a_nr^n+\frac{b_1}{r}+\frac{b_2}{r^2} \nonumber \\
&& +\cdots+\frac{b_m}{r^m}) f_2(\theta), \nonumber \\
g^{\theta\theta} &=& f_3(r), \nonumber
\end{eqnarray}
where $f_1$,  $f_2$ and  $f_3$ are arbitrary continuous
differentiable functions when $r$ is restricted to $r>2$, and
$a_0$, $\cdots$,  $a_n$, $b_1$, $\cdots$, $b_m$ are constant
parameters. For the Kerr metric, the operator splitting method is
not available because the denominators of $g^{rr}$ and
$g^{\theta\theta}$ are $r^2+a^2\cos^2\theta$. Similarly, the
method becomes useless for the magnetized Ernst-Schwarzschild
spacetime (Ernst 1976).

\subsection{Numerical evaluations}

Now, let us apply the second-order explicit symplectic  method S2
or the fourth-order method S4 to work out the magnetized RN-dS
spacetime (11). We take the parameters $E=0.975$, $L=4.2$,
$\beta=8\times10^{-3}$, $Q=0.3$, $Q^{*}=1\times 10^{-6}$, and
$\Lambda=2\times 10^{-6}$. A test orbit has initial conditions
$r=25$, $p_r=0$ and $\theta=\pi/2$. The starting value of
$p_{\theta}>0$ is determined by Eq. (14). Proper time step $h=1$
is adopted. It can be seen clearly from  Fig. 1 (a) that algorithm
S2 performs bounded Hamiltonian errors $\Delta K=-1/2-K$ with an
order of $\mathcal{O}(10^{-6})$ in a long numerical integration of
$10^9$ steps. The errors of S4 are smaller in 3 orders than those
of S2, but have a slightly secular drift due to roundoff errors.
Fortunately, this secular error growth is absent when a larger
proper time step $h=4$ is used in the fourth-order method S4*. The
fourth-order algorithm with the larger time step $h=4$ is the same
as the second-order method with the smaller time step $h=1$ in the
order of Hamiltonian errors. The former efficiency is 4/3 times
faster than the latter one.

In fact, the test orbit is a regular orbit colored red in Fig. 1
(b). This regular orbit seems to consist of 7 loops on
Poincar\'{e} section at the plane $\theta=\pi/2$ and
$p_{\theta}>0$ (note that the 7 loops are unclearly visible). The
order of Hamiltonian errors for each algorithm is not altered when
the red regular orbit is replaced with the green regular orbit,
blue regular orbit, or black weak chaotic orbit in Fig. 1 (b). In
other words, the algorithms' accuracies are independent of
dynamical behavior of orbits. In what follows, we use the
second-order method S2 with the smaller time step $h=1$ to trace
the dynamics of order and chaos of orbits.

\subsection{Dynamical transition with a  variation of the cosmological
constant}

To show cosmological constant $\Lambda$ how to affect the orbital
dynamics of the system (11), we consider $\Lambda$ to increase.
The red orbit is regular for $\Lambda=2\times 10^{-6}$ in Fig. 1
(b), but chaotic for $\Lambda=2.1\times 10^{-6}$ in Fig. 2 (a).
The extent of chaos  in Figs. 2 (b) and (c) is further
strengthened with an increase of positive cosmological constant
$\Lambda$. In fact, strong chaos occurs for
$\Lambda=1\times10^{-5}$. However, no chaos exists for $\Lambda=0$
in Fig. 2 (d). In particular, the red orbit is clearly composed of
7 small islands and is approximately a periodic orbit in this
case.  These results seem to show that chaos gets stronger as a
positive cosmological constant increases. For negative
cosmological constants, such as $\Lambda=-5\times10^{-6}$ and
$\Lambda=-1\times10^{-5}$ in Figs. 2 (e) and (f), chaos is absent,
either.

The magnetic parameter $\beta=8\times10^{-3}$ is taken in Figs. 1
and 2. What about the dynamics of the system (11) when $\beta$
gets slightly larger? For $\beta=9\times10^{-3}$ with  $\Lambda=0$
in Fig. 3 (a), two strong chaotic orbits exist, but do not for
$\beta=8\times10^{-3}$ with  $\Lambda=0$ in Fig. 2 (d). As the
positive cosmological constant further increases, chaos is easier
to occur. For $\Lambda=2\times10^{-6}$ in Fig. 3 (b), three
chaotic orbits are present.  For $\Lambda=1\times10^{-5}$ in Fig.
3 (c), the four orbits are chaotic. However, an increase of the
magnitude of negative  cosmological constant weakens the extent of
chaos. For example, two weak chaotic orbits exist for
$\Lambda=-1.2\times10^{-6}$ in Fig. 3 (e). When
$\Lambda=-1.5\times10^{-6}$ in Fig. 3 (f), no orbits can be
chaotic.

The aforementioned demonstrations are based on the choice of a
small value of the Coulomb parameter $Q^{*}$. Now, let us focus on
the orbital dynamical transition when larger values of $Q^{*}$ are
considered. Comparisons between Figs. 4 (a) and (b), Figs. 4 (c)
and (d), Figs. 4 (e) and (f), Figs. 5 (g) and (h), and Figs. 5 (i)
and (j) still consistently support the result that chaos becomes
easier with an increase of positive cosmological constant
$\Lambda$. In particular, this result is irrespective of whether
$Q$ and $Q^{*}$ are larger or smaller, and  $Q^{*}$ is positive or
negative. However, an increase of the magnitude of negative
cosmological constant leads to suppressing the occurrence of chaos
or weakening the extent of chaos, as shown by the comparison
between Figs. 5 (k) and (l). The chaos weakened is qualitatively
observed from the Poincar\'{e} section. The blue orbit is chaotic
for $\Lambda=-2\times10^{-6}$ in panel (k), but it is not for
$\Lambda=-1\times10^{-5}$ in panel (l). In addition, other results
are visible in Figs. 4 and 5. Figs. 1 (b), 2 (c), and 4 (a) and
(b) describe that chaos is gradually weakened as a positive
Coulomb parameter $Q^{*}$ increases. And chaos is much easily
induced when the magnitude of negative Coulomb parameter
increases. See Figs. 4 (e) and (f), and 5 (g) and (h) for more
details. These results are consistent with those of Paper II (Wang
et al. 2021b). As another one of the main results in Paper II,  an
increase of $Q$ does not exert a typical influence on the
occurrence of chaos but twists the shape of orbits (such as the
green orbit). This result is also confirmed by the comparisons
among Figs. 4 (a)-(d), and 5 (g)-(j).

It can be concluded from Figs. 1 (b), 2, 3, 4 and 5 that an
increase of positive cosmological constant (or an increase of the
magnitude of negative Coulomb parameter $Q^{*}$) leads to
strengthening the extent of chaos from the global phase space.
However, an increase of the magnitude of negative cosmological
constant (or an increase of positive Coulomb parameter $Q^{*}$)
does not. These results can be explained similarly with the aid of
slightly modified Equation (36) in Paper II:
\begin{eqnarray}
K_1 &\approx& -\frac{1}{2}(\beta
L+E^2)-\frac{\Lambda}{6}Q^{*2} -\frac{1}{6}\Lambda E^2 r^{2} \nonumber \\
&&  +\frac{1}{3}\Lambda EQ^{*} r-\frac{E^2}{r}+\frac{\beta^2}{8} r^{2}\sin^{2} \theta  \nonumber \\
&& +\frac{EQ^{*}}{r} +\frac{Q^2E^2}{2r^2} + \frac{L^2}{2r^2\sin^2\theta} \nonumber \\
&& +\frac{Q^{*}}{2r^2}(4E-Q^{*})+\cdots.
\end{eqnarray}
The third and fourth terms in the  potential (31) correspond to
the forces associated with the cosmological constant. The third
term denominates the fourth term. It acts as  a repulsive force
for the RN-dS black hole. The repulsive force reduces the
attractive force given by the potential $-E^2/r$ of the black
hole. However, the third term yields an attractive force for the
RN-AdS black hole. The attractive force enhances the attractive
force from the black hole. On the other hand, the potential
$EQ^{*}/r$ corresponds to the Coulomb force from the Coulomb part
of the magnetic field potential. For $Q^{*}>0$,  the Coulomb force
is a repulsive force, which weakens the Lorentz force as an
attractive force determined by the potential $\beta^2
r^{2}\sin^{2} \theta/8$ of the magnetic field. However, the
Coulomb force is an attractive force for $Q^{*}<0$. This leads to
enhancing the attractive force from the magnetic field. Clearly,
the Coulomb force from the Coulomb part of the magnetic field
potential is larger than the repulsive force governed by the
potential $Q^2E^2/(2r^2)$ from the black hole's charge. Only when
the gravitational attractions of charged particles from the black
hole and the magnetic field are approximately balanced, may chaos
occur. Therefore,  chaos is easily caused  in some circumstances
when any one of the positive cosmological constant, the magnitude
of negative Coulomb parameter, and the magnetic field parameter
increases.

\section{Conclusion}

The Hamiltonian for the description of charged particles moving
around the RN-(A)dS black hole with an external magnetic field can
be split into six parts, which have  analytical solutions as
explicit functions of proper time. In this case, second- and
fourth-order explicit symplectic integration methods are easily
designed for this Hamiltonian system.

It is shown via numerical simulations that the newly proposed
algorithms with appropriate choices of step-sizes perform good
long-term performance in maintaining the boundness of Hamiltonian
errors. The result should always be the same, regardless of
whether a test orbit is regular or chaotic.

The cosmological constant in the RN-dS black hole acts as a
repulsive force, which weakens the gravitational force of charged
particle from the black hole. As a result, the extent of chaos is
strengthened from  the global phase space when a positive
cosmological constant increases. In other words, the accelerated
expansion of the universe can easily induce chaos of charged
particles in some cases. However, chaos is weakened when the
magnitude of negative cosmological constant increases. This is
because the cosmological constant in the RN-AdS black hole acts as
an attractive force, which causes  the gravitational effect from
the black hole to dominate the attractive force from the magnetic
field. For the presence of chaos in the present metric background,
the effect of the cosmological constant on the accretion rate or
the gravitational waves will be worth considering.

\section*{Acknowledgments}

The authors are very grateful to a referee for useful suggestions.
This research has been supported by the National Natural Science
Foundation of China [Grant Nos. 11973020 (C0035736), 11533004,
11803020, 41807437, U2031145], and the Natural Science Foundation
of Guangxi (Grant Nos. 2018GXNSFGA281007 and 2019JJD110006).

\begin{figure*}[ptb]
\center{
\includegraphics[scale=0.25]{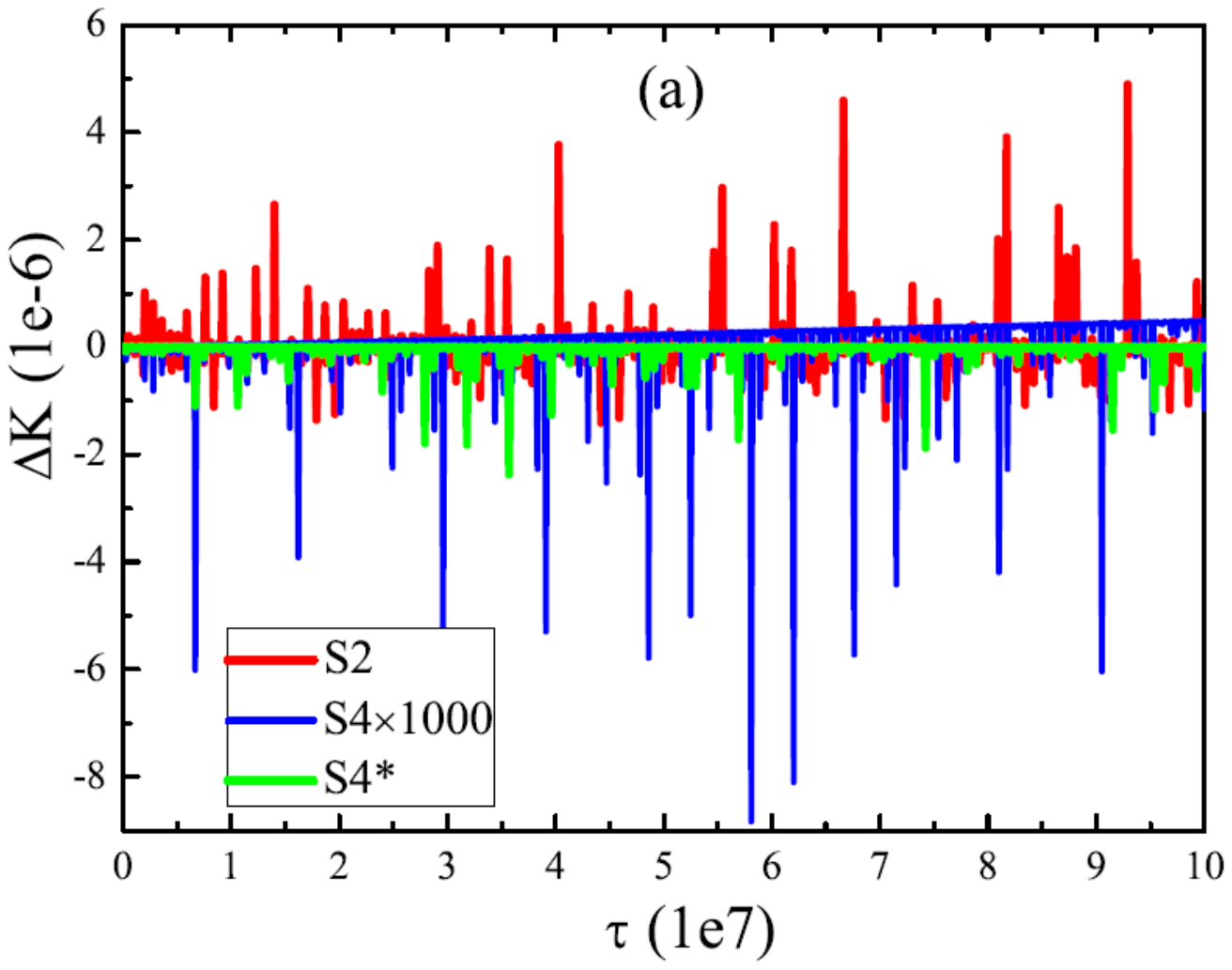}
\includegraphics[scale=0.25]{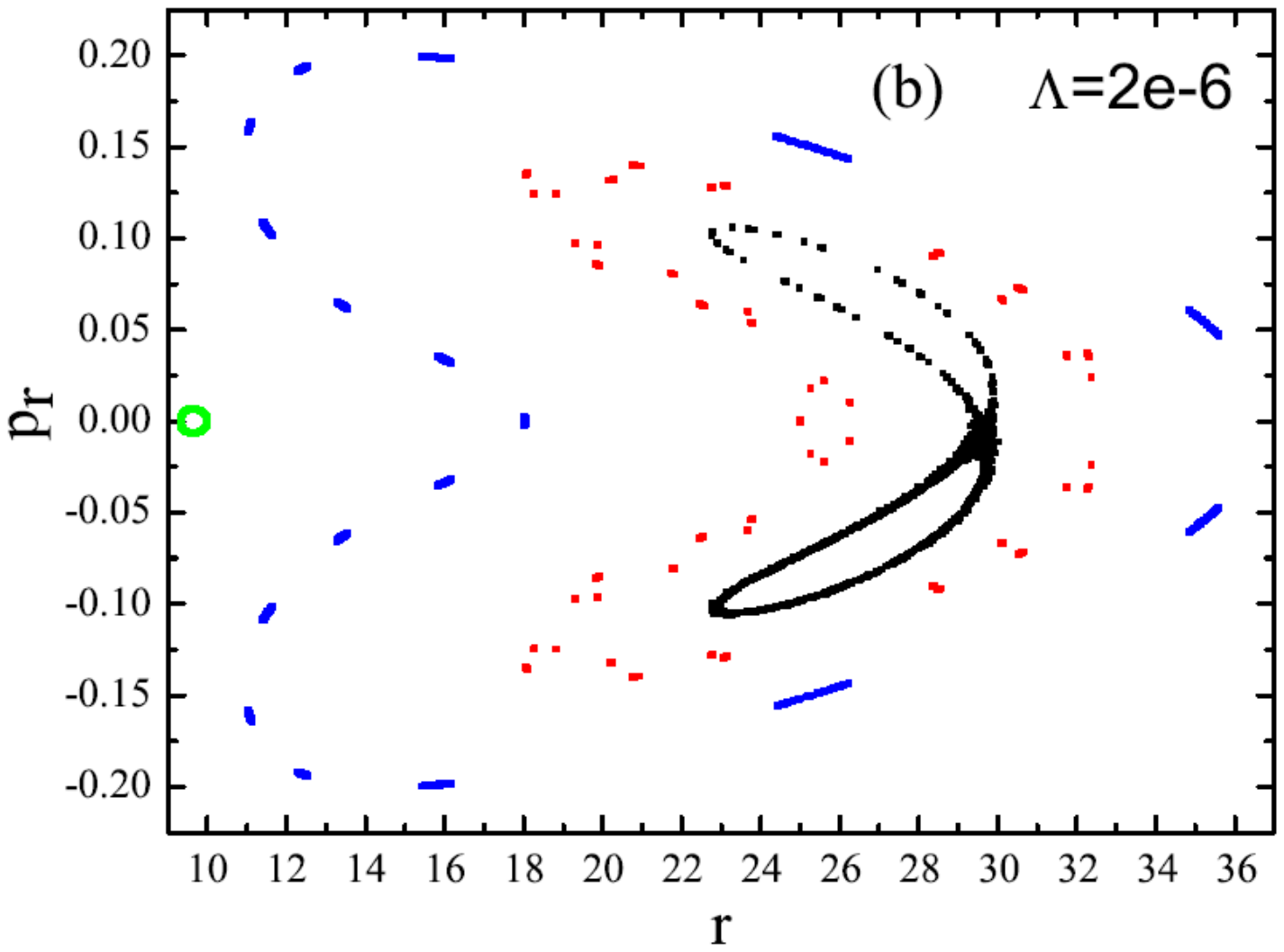}
\caption{(a) Hamiltonian errors $\Delta K=-1/2-K$ for the new
explicit symplectic algorithms solving the magnetized RN-dS
spacetime (11). The parameters $E=0.975$, $L=4.2$,
$\Lambda=2\times10^{-6}$, $\beta=8\times10^{-3}$, $Q=0.3$ and
$Q^{*}=1\times10^{-6}$ are used, and the initial conditions of a
test orbit are $r=25$, $p_r=0$ and $\theta=\pi/2$ with the
starting value of $p_{\theta}>0$ determined by Eq. (14) are
chosen. The second-order method S2 and the fourth-order method S4
take proper time step $h=1$. The realistic errors for S4 are 1000
times smaller than the plotted errors. The errors for S2 are
stabilized in an order of $\mathcal{O}(10^{-6})$, whereas do not
remain bounded for S4 due to roundoff errors. Fortunately, such a
secular drift in Hamiltonian errors lose when a larger time step
$h=4$ is used in method S4*. (b) Poincar\'{e} sections on the
plane $\theta=\pi/2$ and $p_{\theta}>0$, given by algorithm S2
with proper time step $h=1$. The test orbit in panel (a) is a
regular orbit colored red in panel (b). The other three orbits are
regular tori colored green with initial separation $r=10$ and blue
with initial separation $r=18$, and a weak chaotic orbit colored
black with initial separation $r=30$.}
 \label{Fig1}}
\end{figure*}

\begin{figure*}[ptb]
\center{
\includegraphics[scale=0.18]{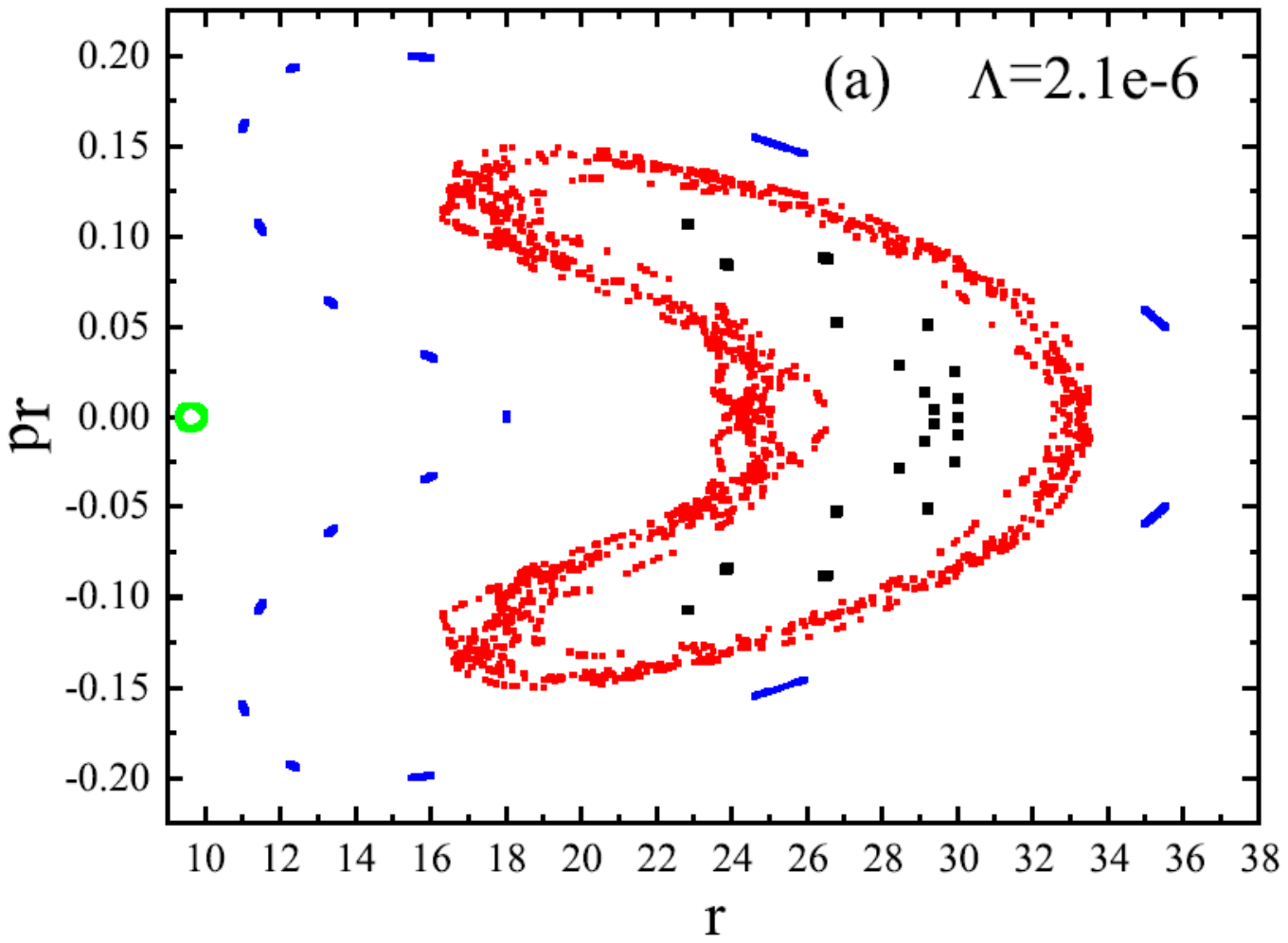}
\includegraphics[scale=0.18]{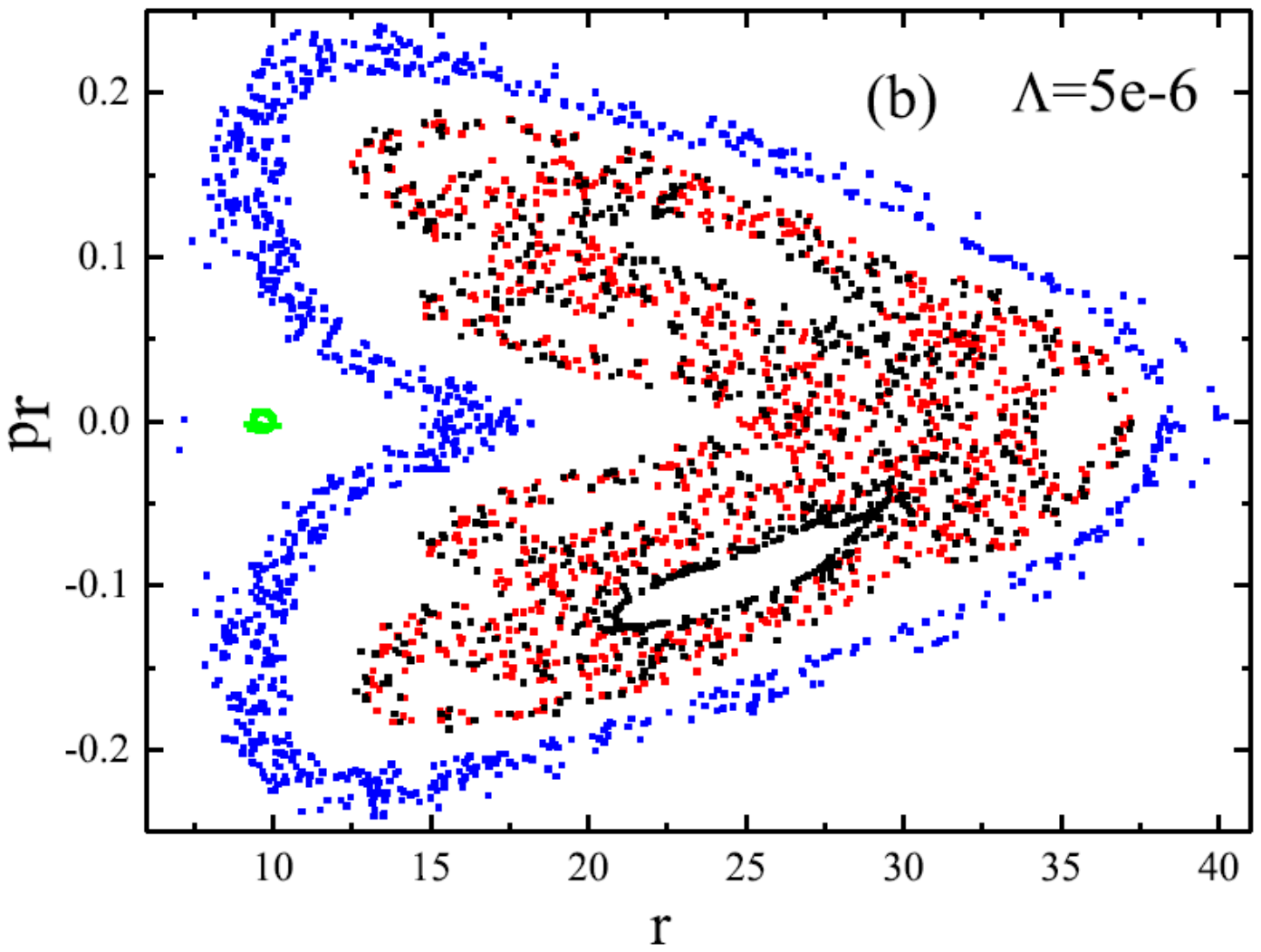}
\includegraphics[scale=0.18]{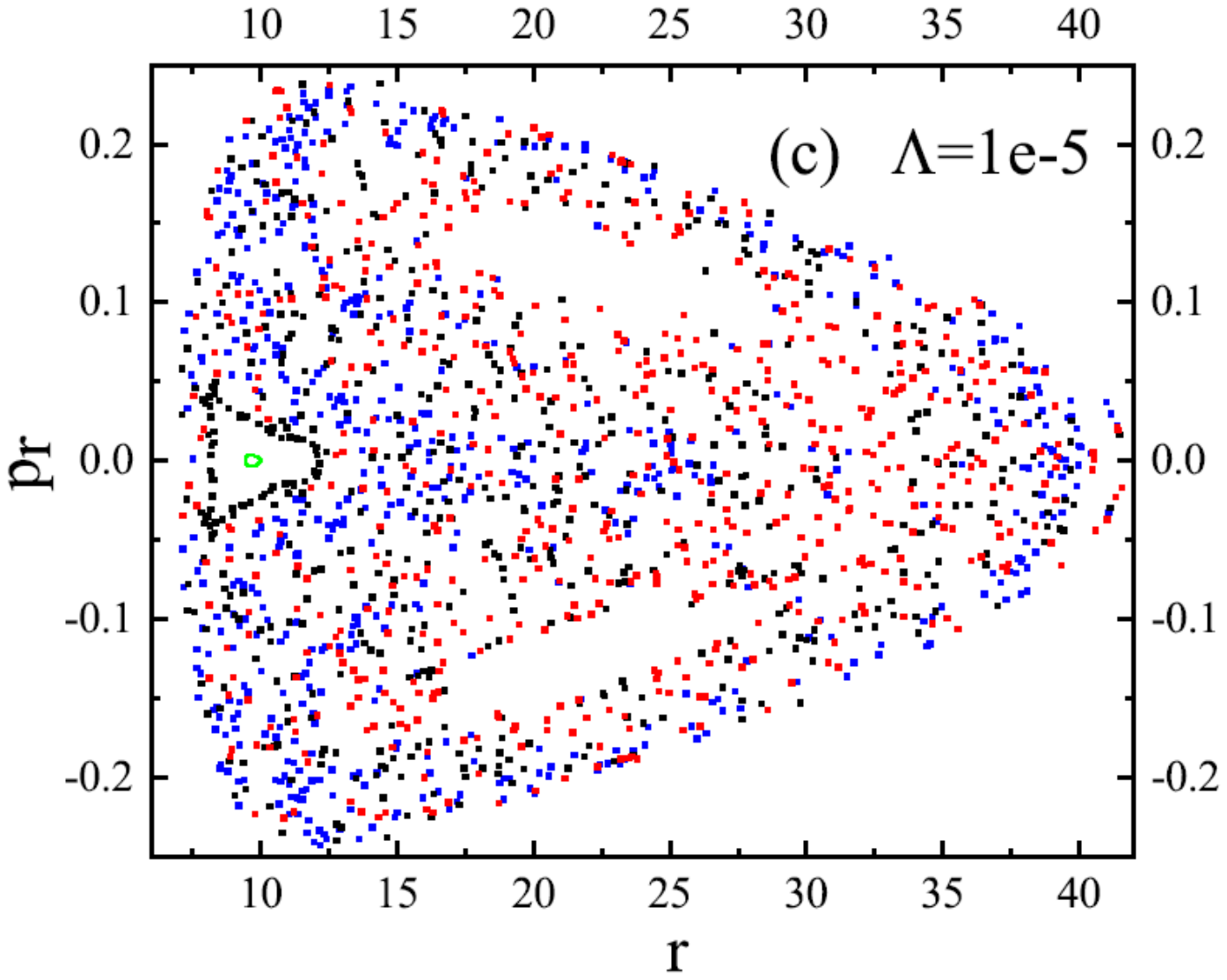}
\includegraphics[scale=0.18]{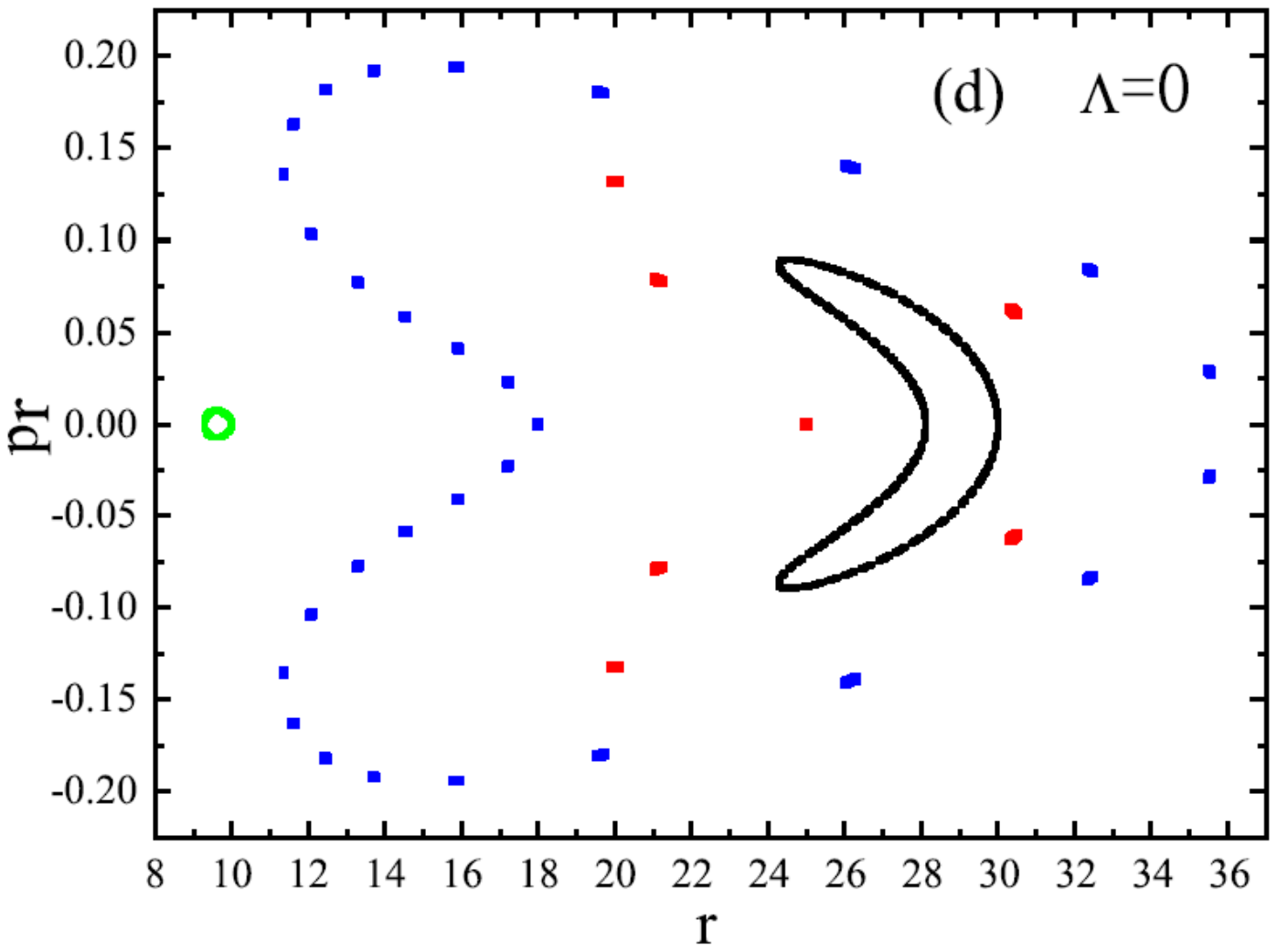}
\includegraphics[scale=0.18]{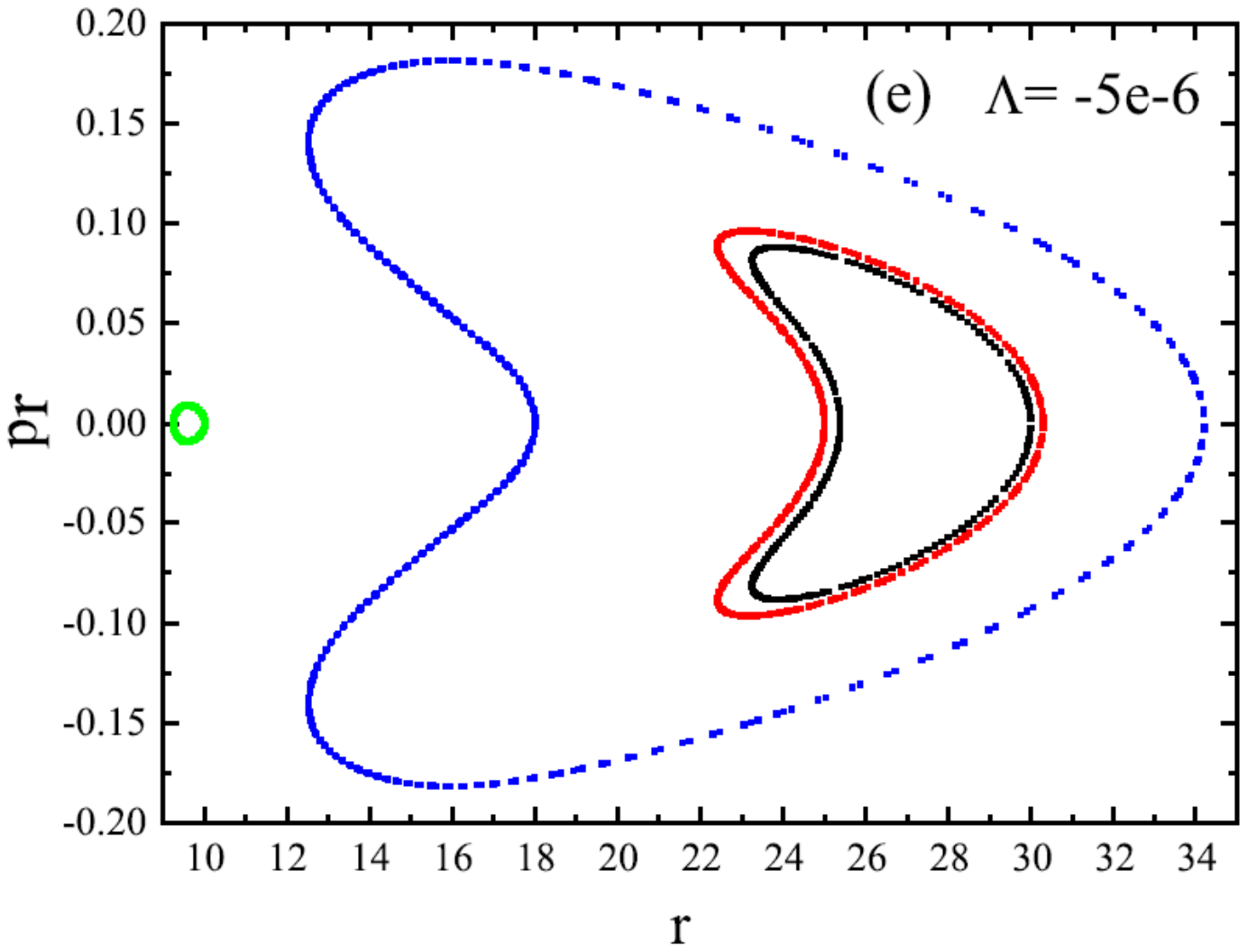}
\includegraphics[scale=0.18]{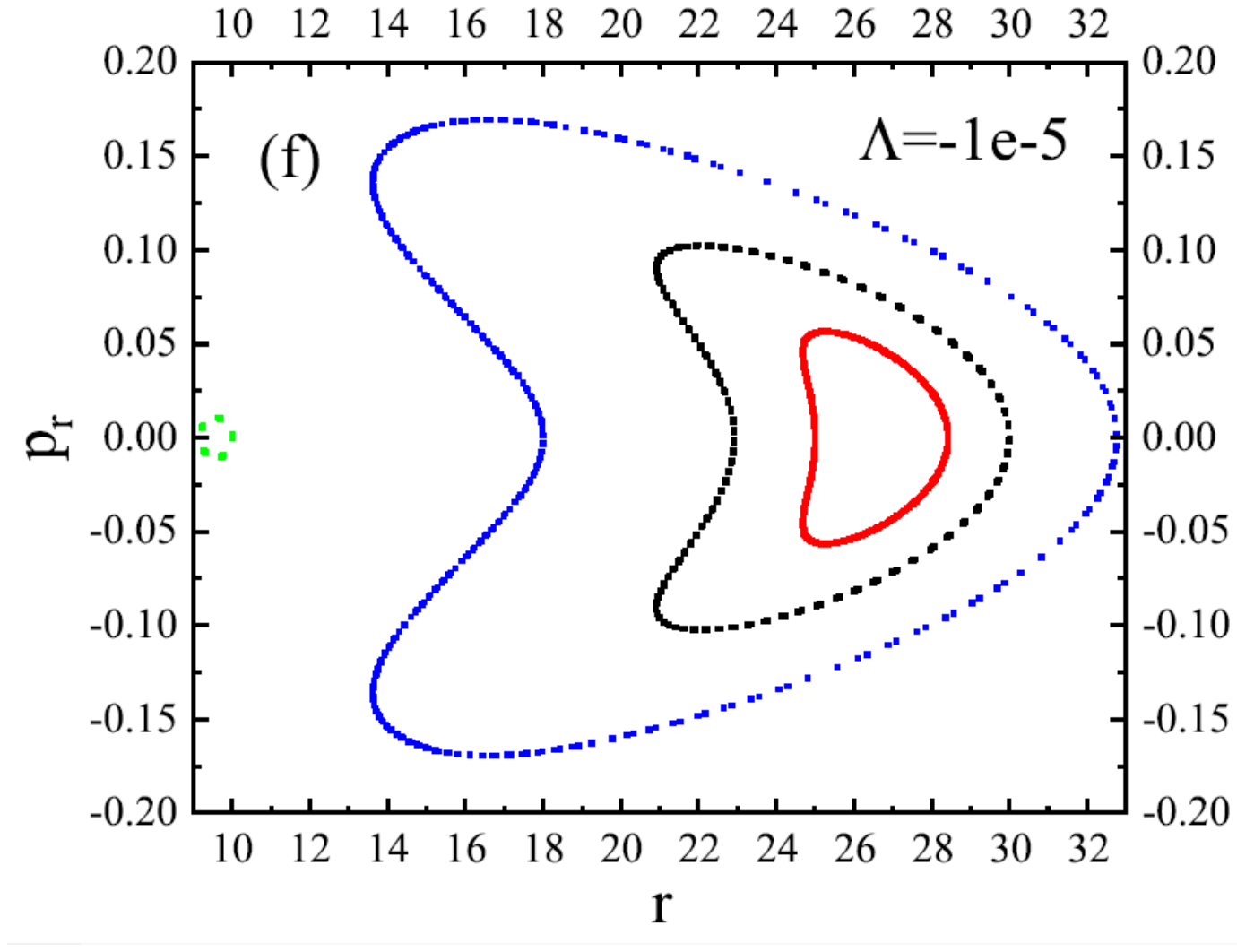}
\caption{Same as Fig. 1 (b) but for different values of
cosmological constant $\Lambda$.}
 \label{Fig2}}
\end{figure*}

\begin{figure*}[ptb]
\center{
\includegraphics[scale=0.18]{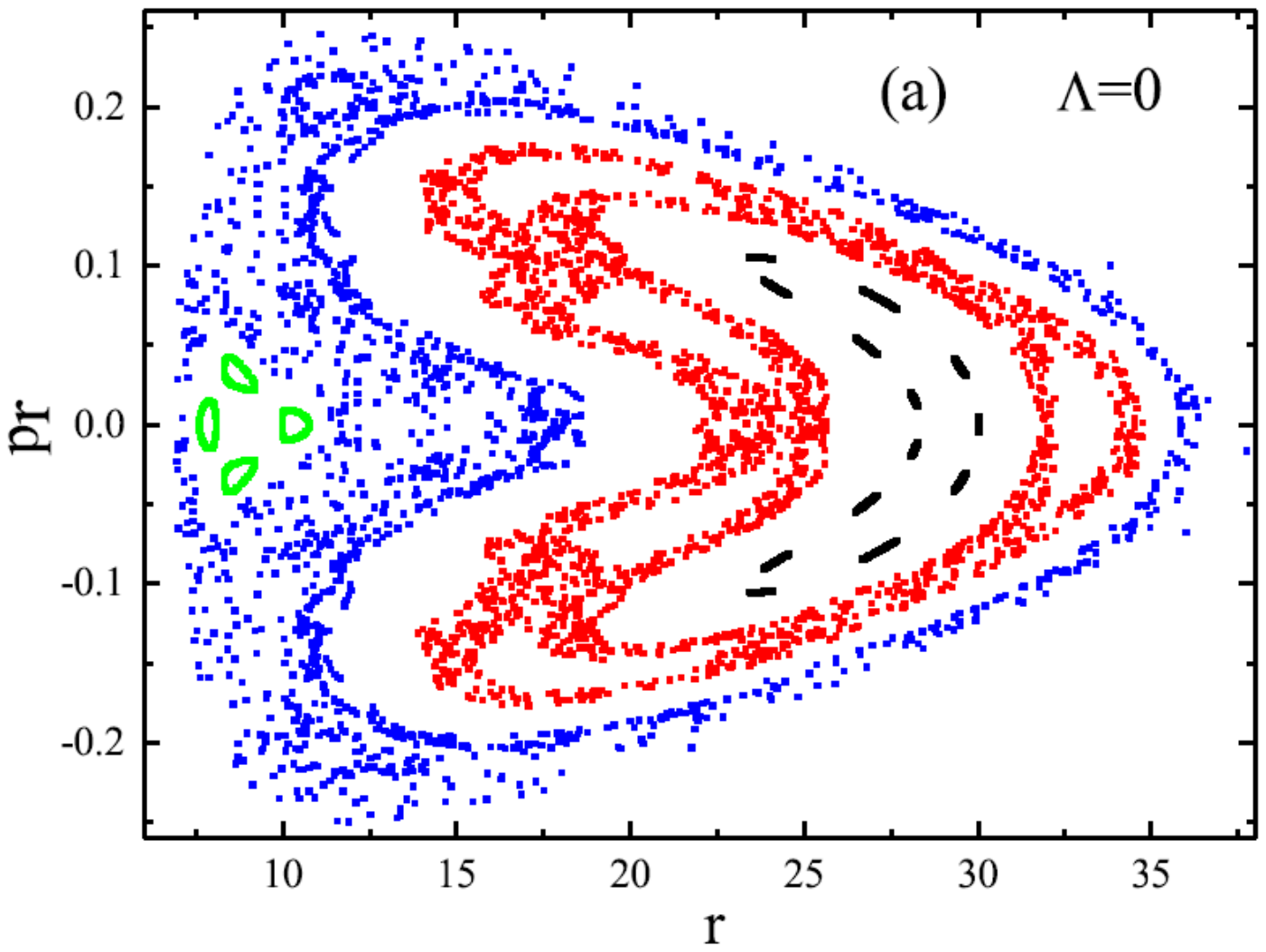}
\includegraphics[scale=0.18]{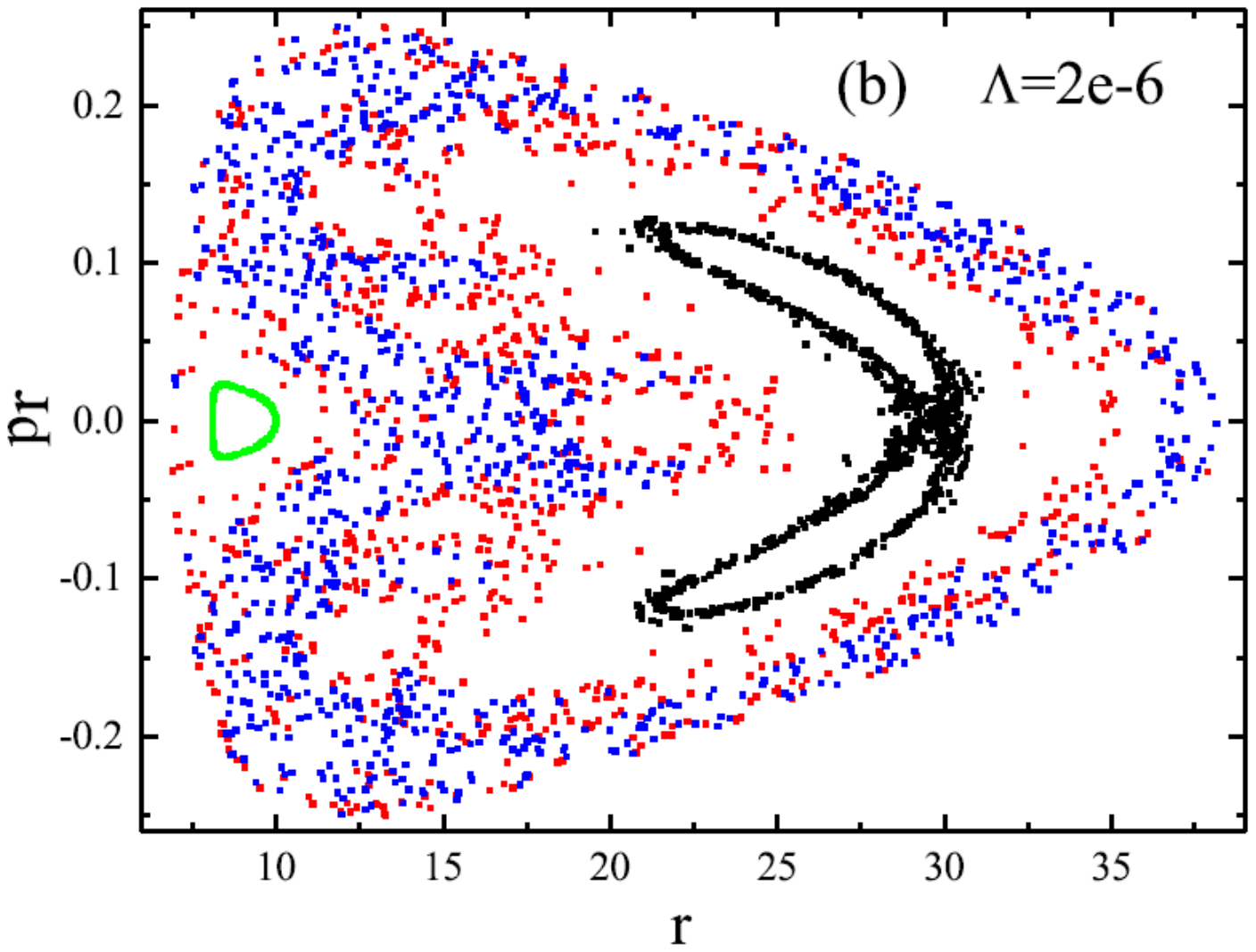}
\includegraphics[scale=0.18]{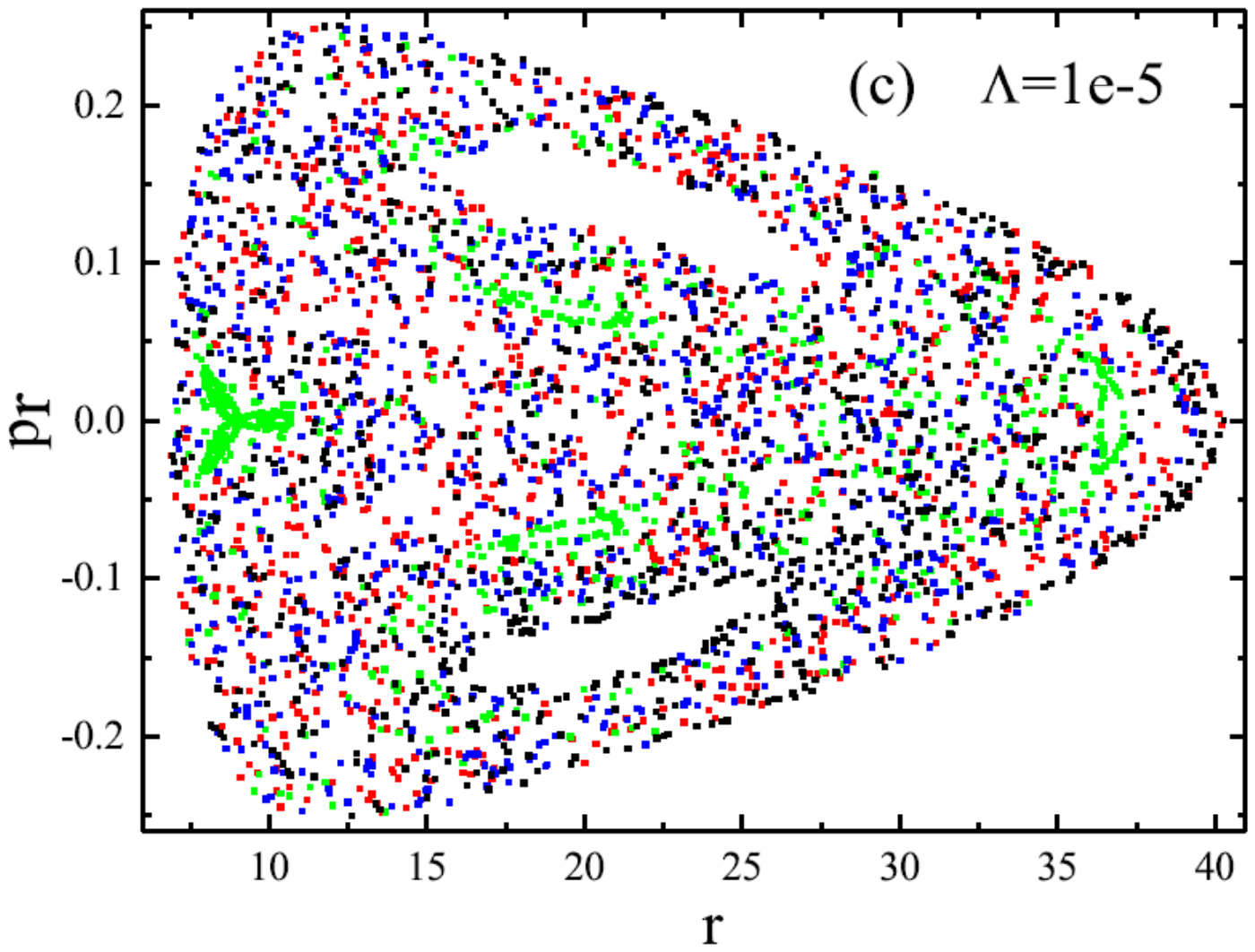}
\includegraphics[scale=0.18]{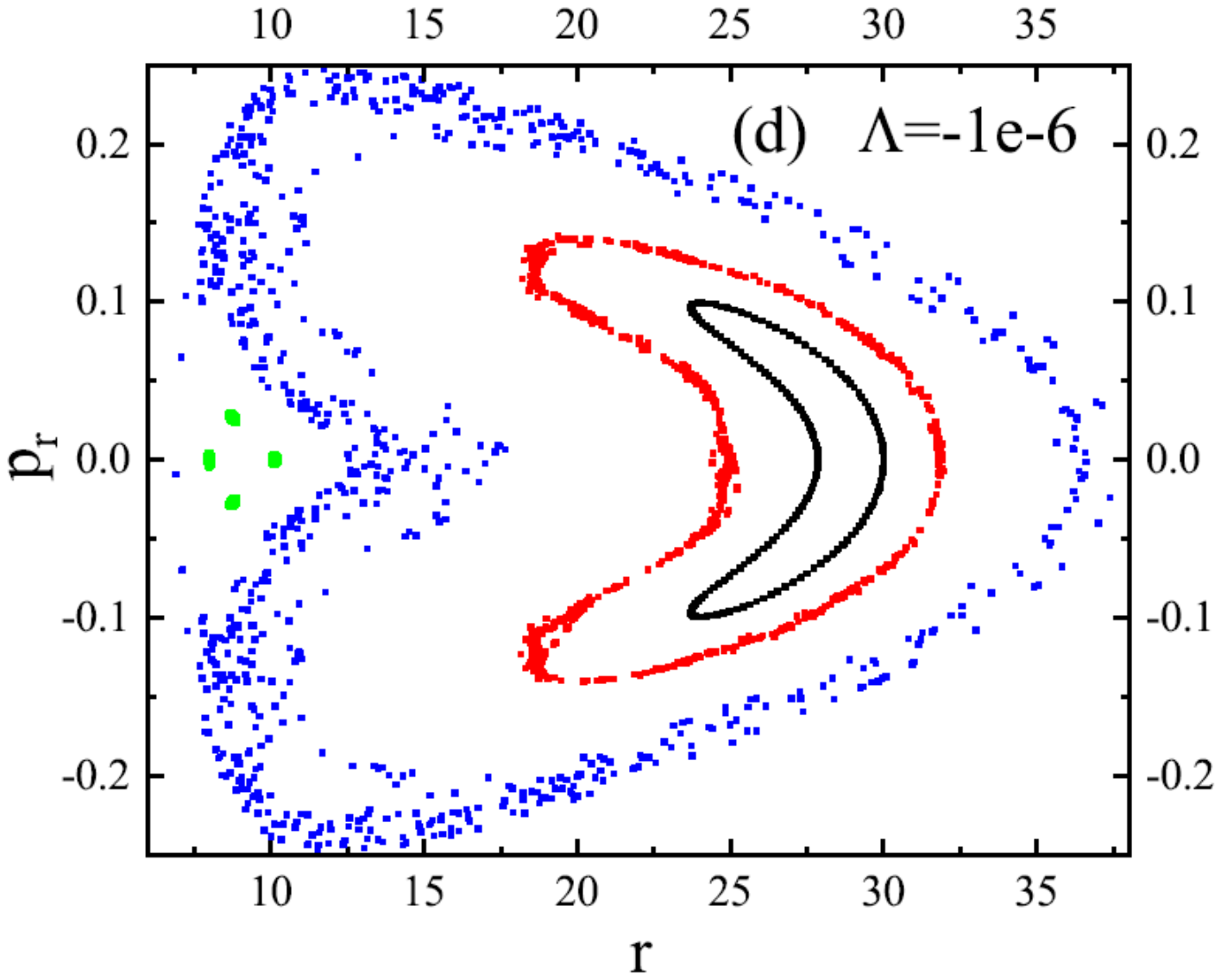}
\includegraphics[scale=0.18]{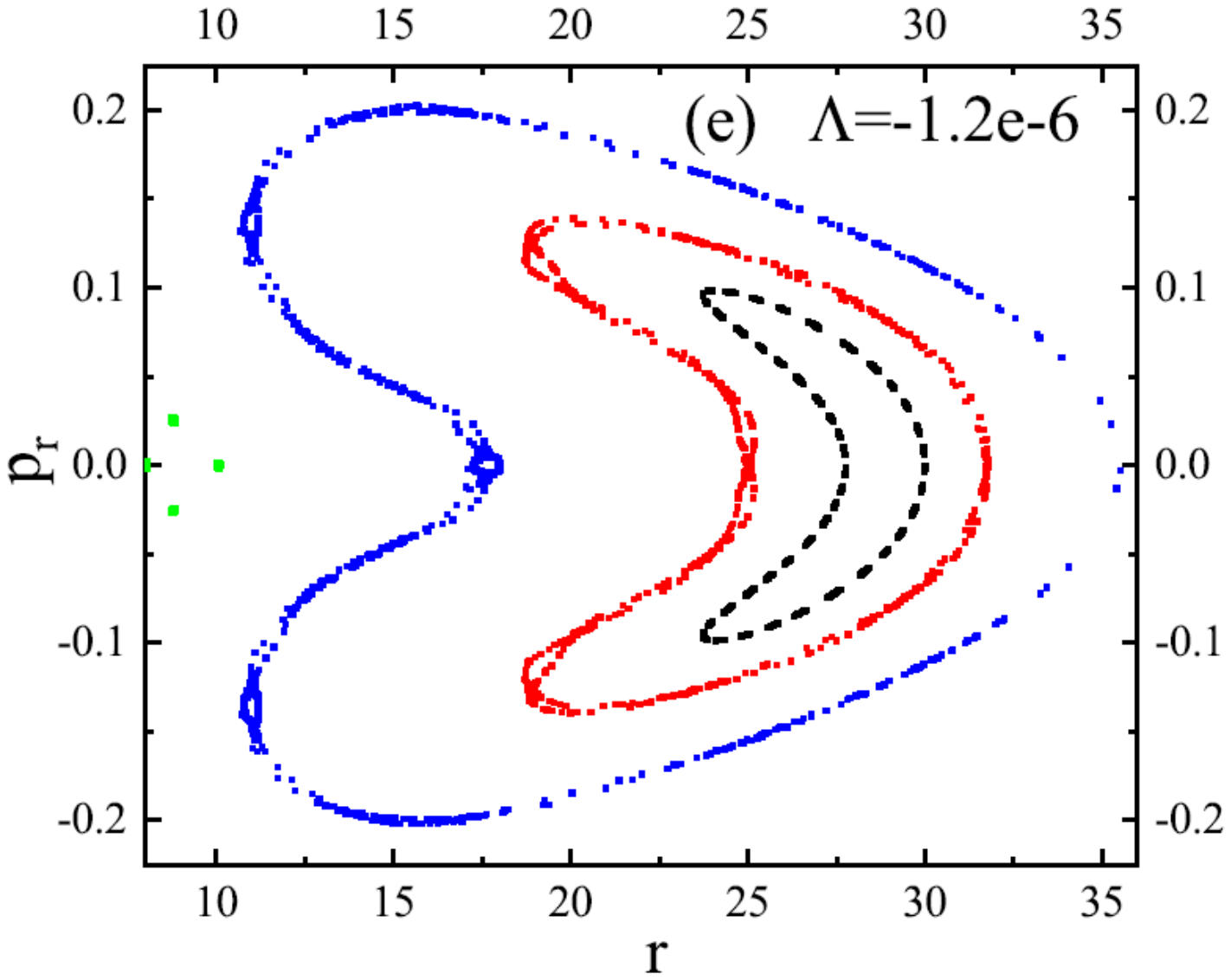}
\includegraphics[scale=0.18]{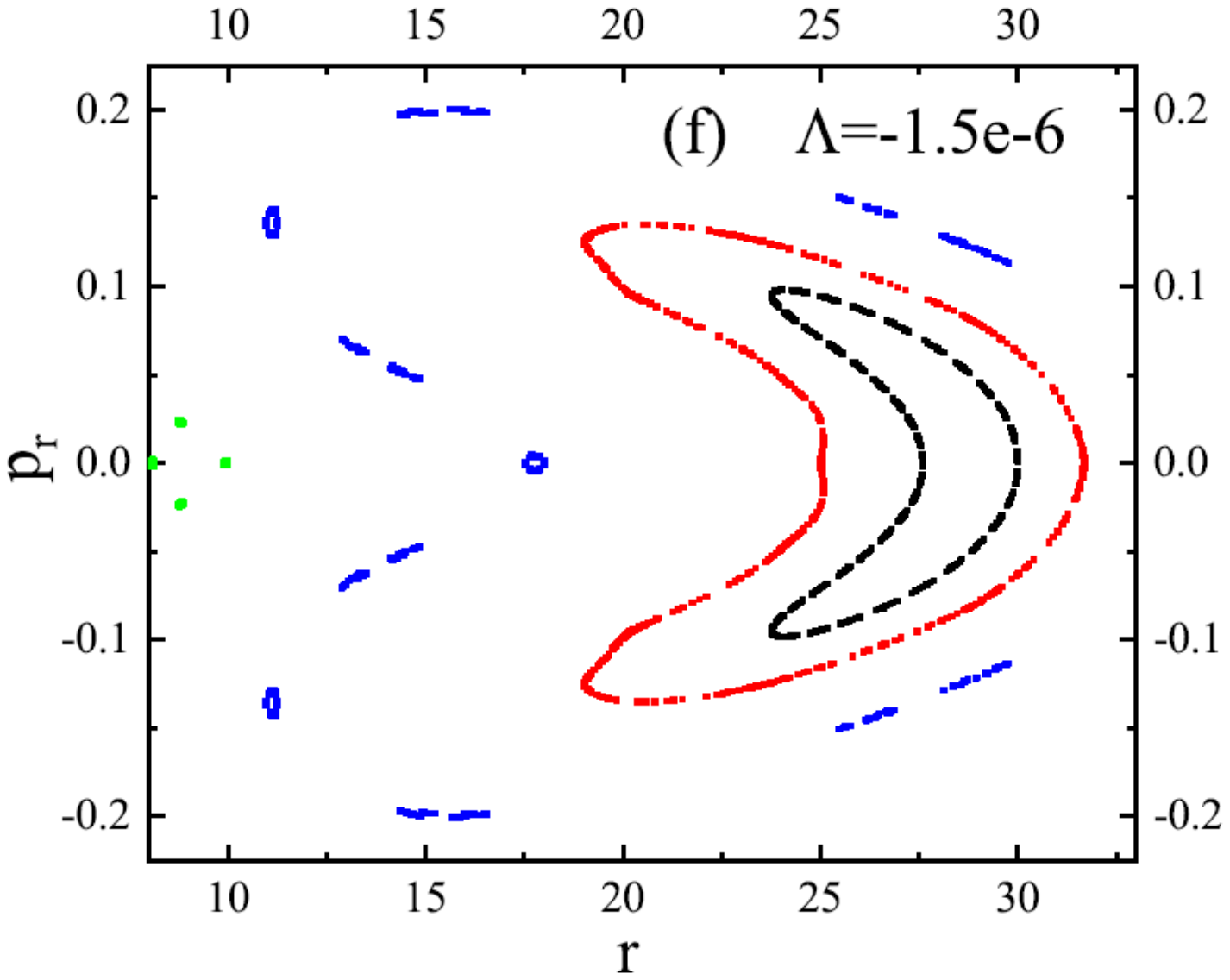}
\caption{Same as Fig. 2 but magnetic parameter
$\beta=9\times10^{-3}$ is fixed.}
 \label{Fig3}}
\end{figure*}

\begin{figure*}[ptb]
\center{
\includegraphics[scale=0.25]{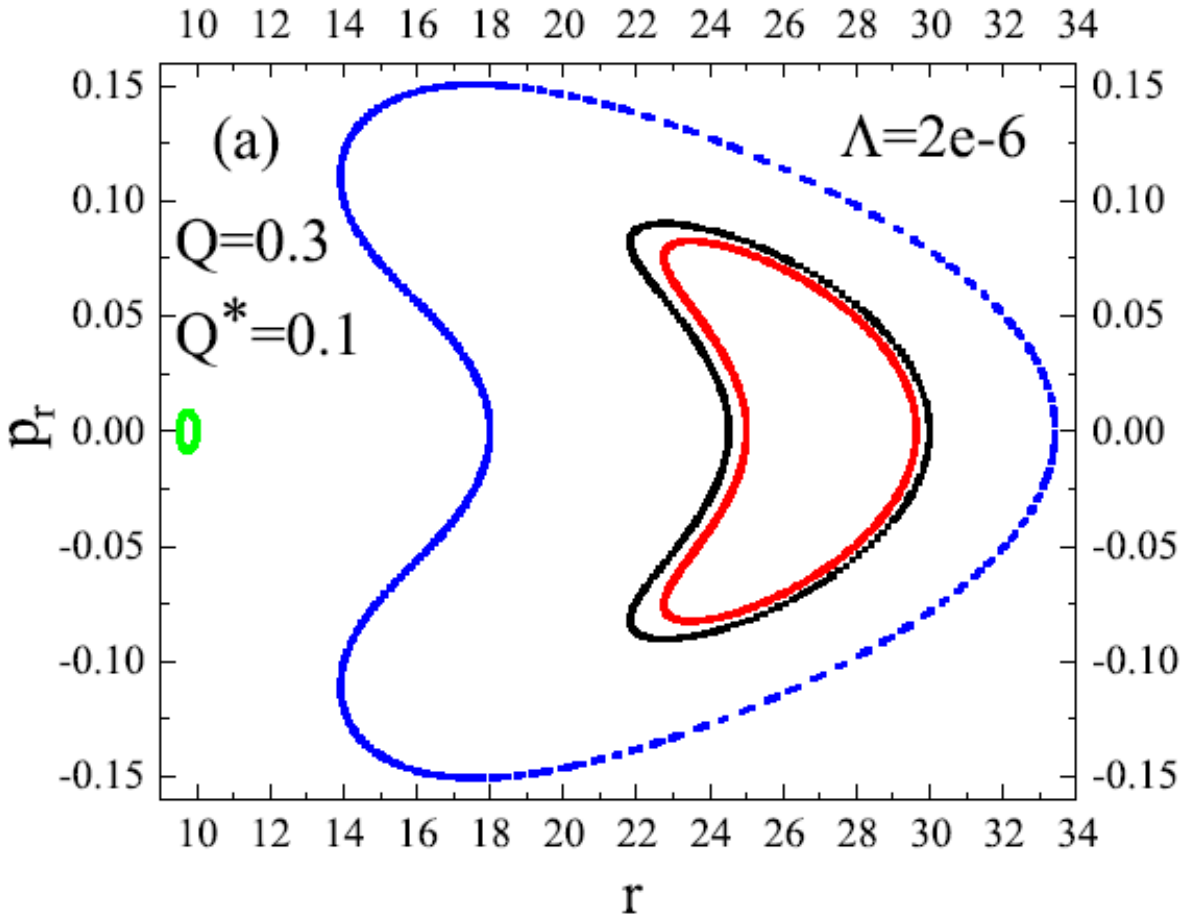}
\includegraphics[scale=0.25]{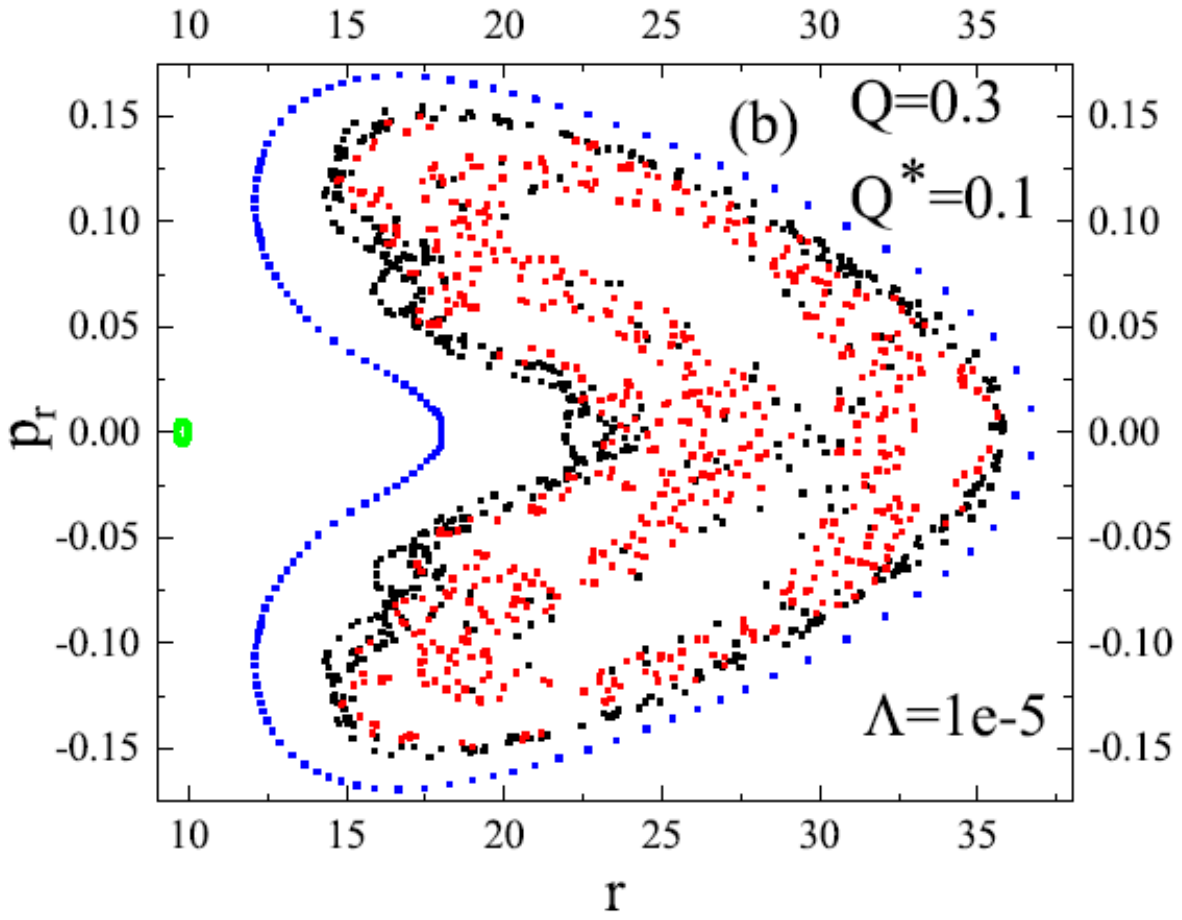}
\includegraphics[scale=0.25]{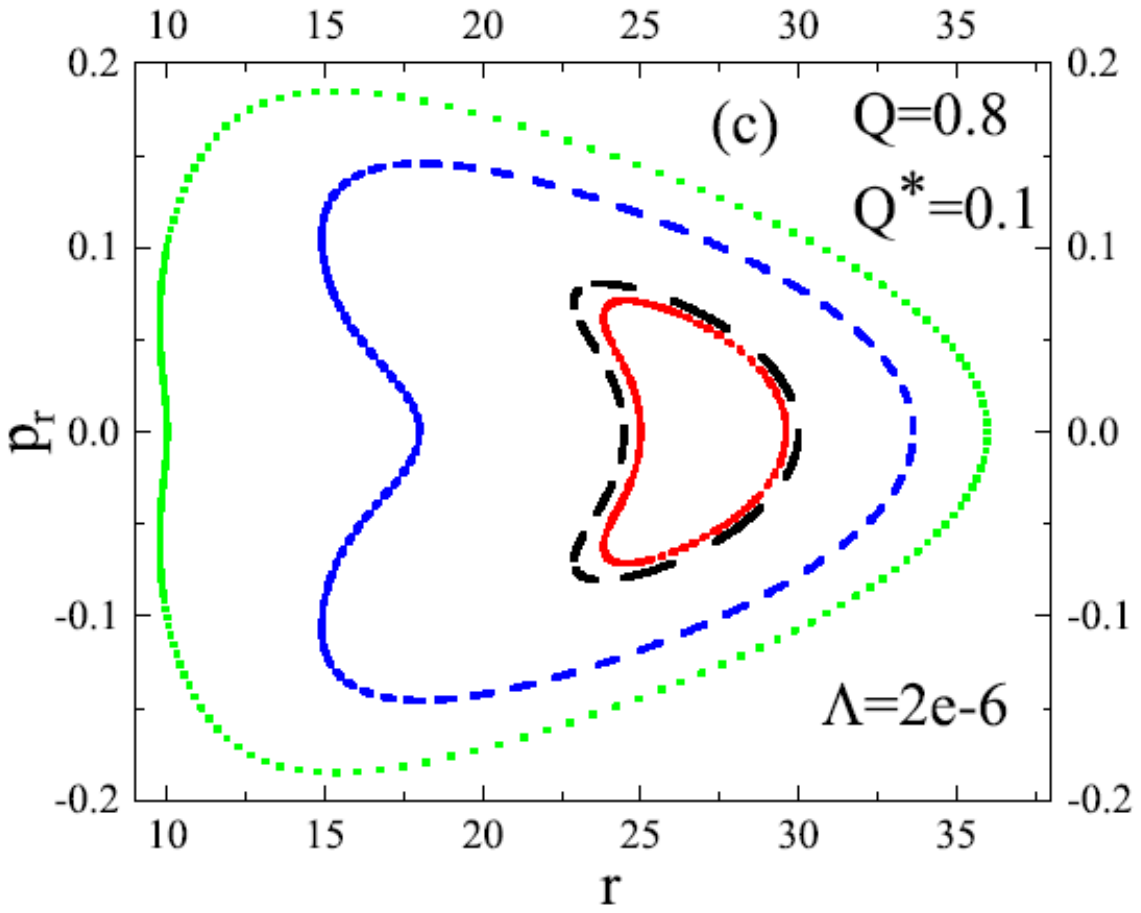}
\includegraphics[scale=0.25]{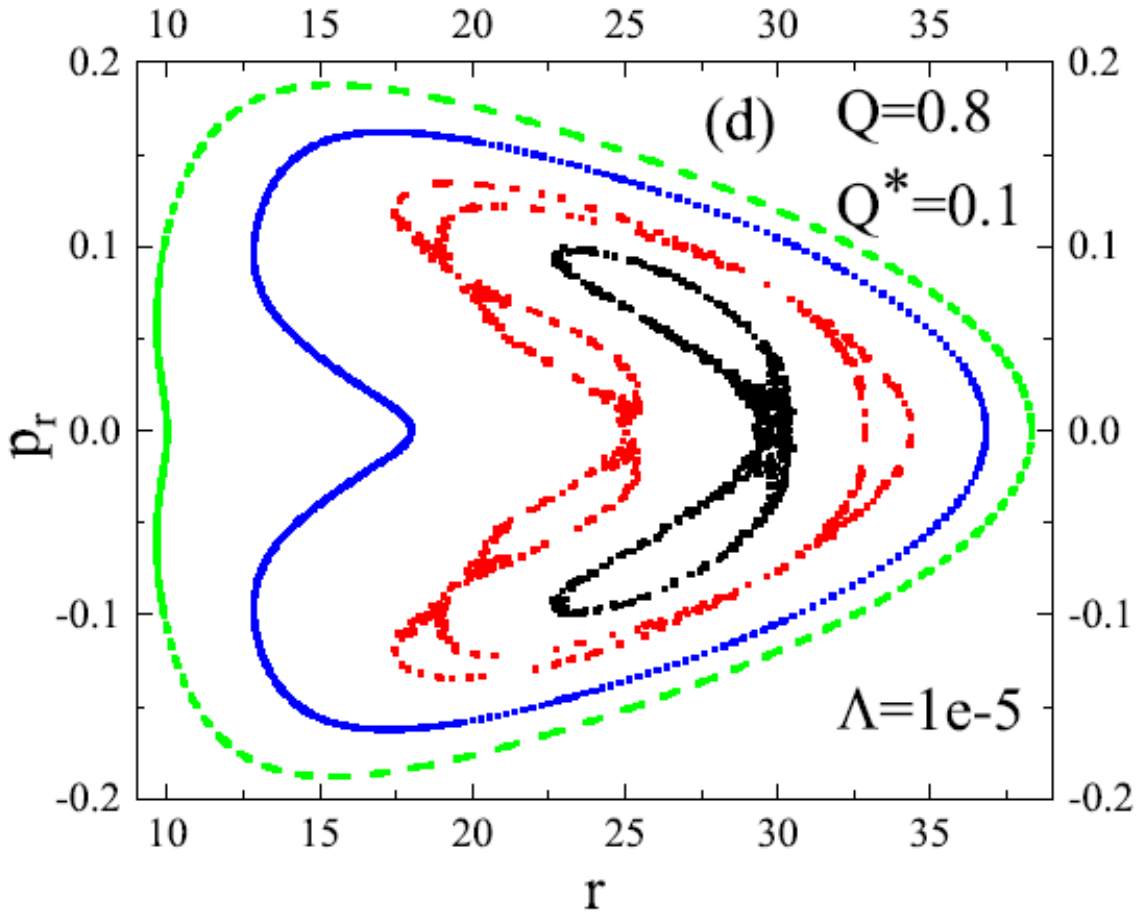}
\includegraphics[scale=0.25]{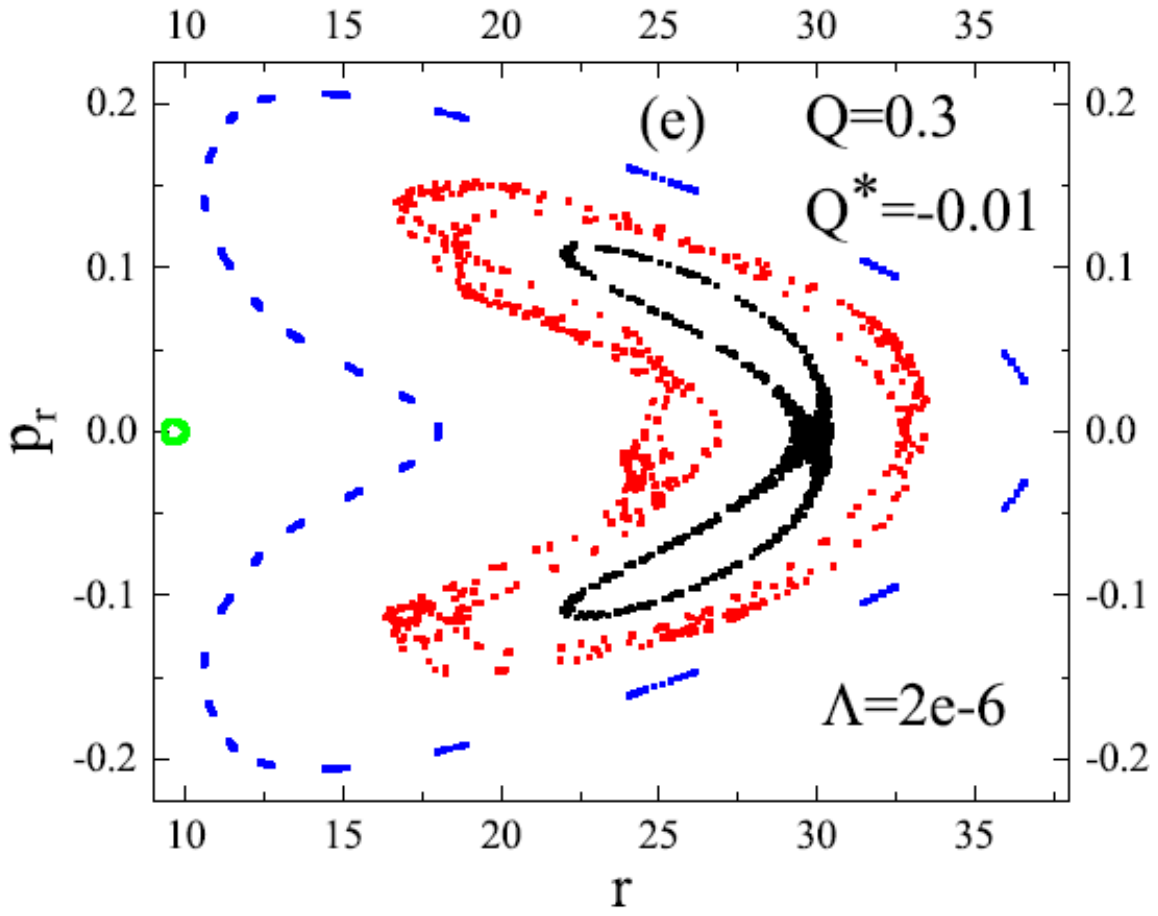}
\includegraphics[scale=0.25]{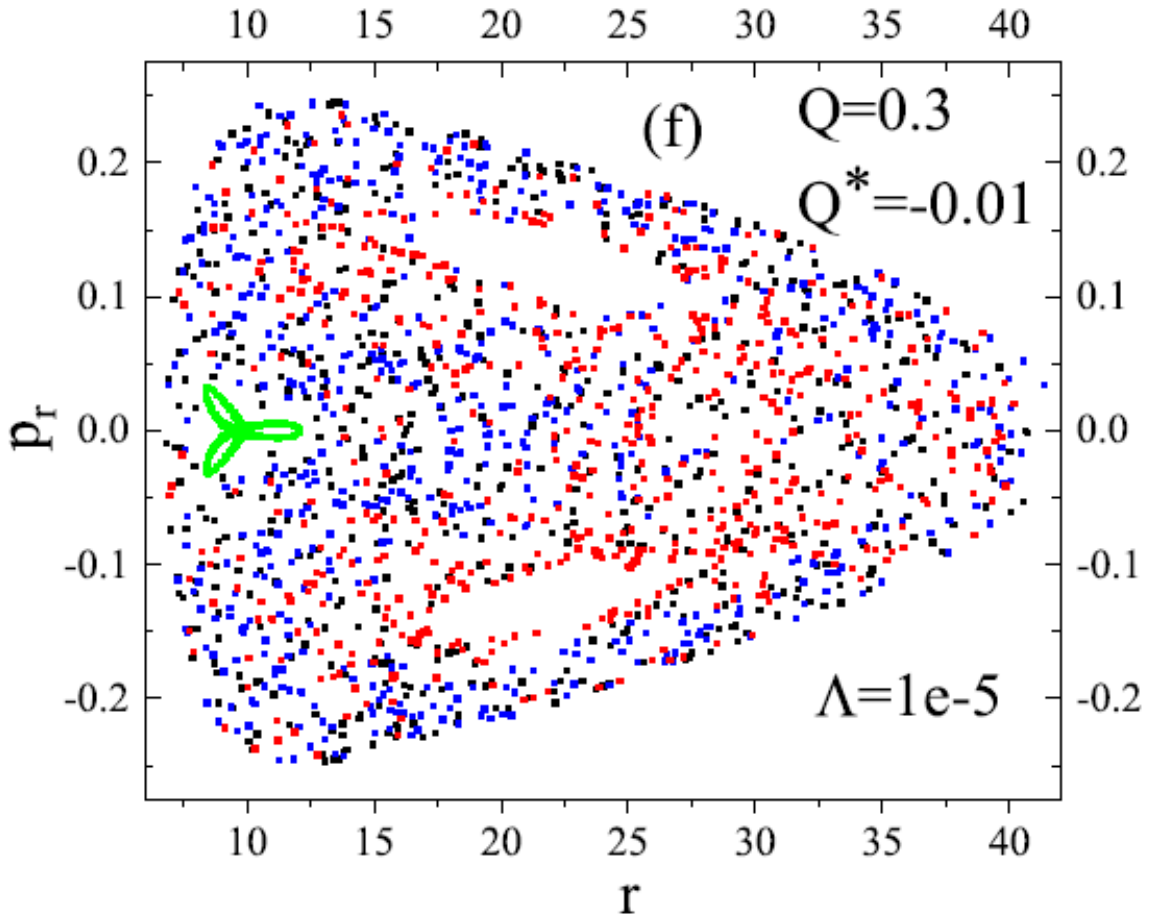}
\caption{Same as Fig. 2 but different combinations of parameters
$Q$, $Q^{*}$ and $\Lambda$ are given.}
 \label{Fig4}}
\end{figure*}

\begin{figure*}[ptb]
\center{
\includegraphics[scale=0.25]{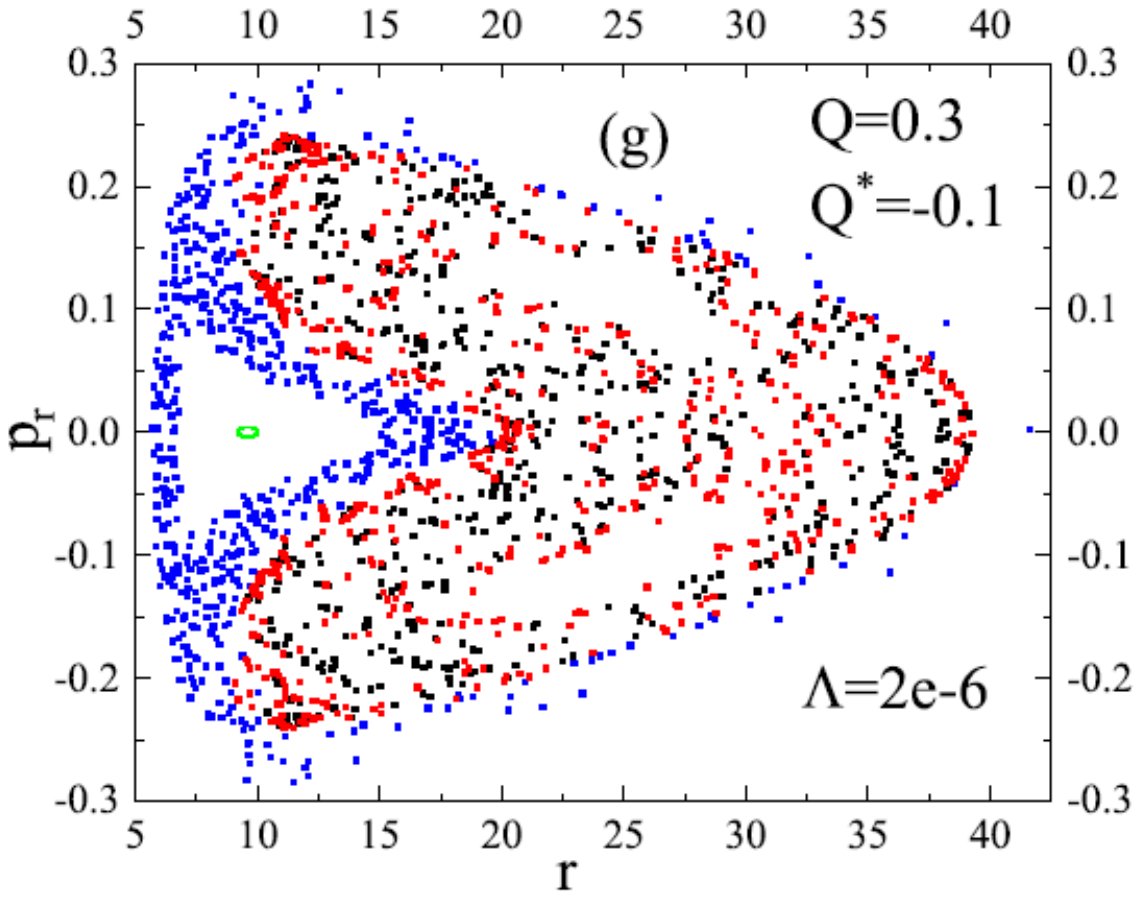}
\includegraphics[scale=0.25]{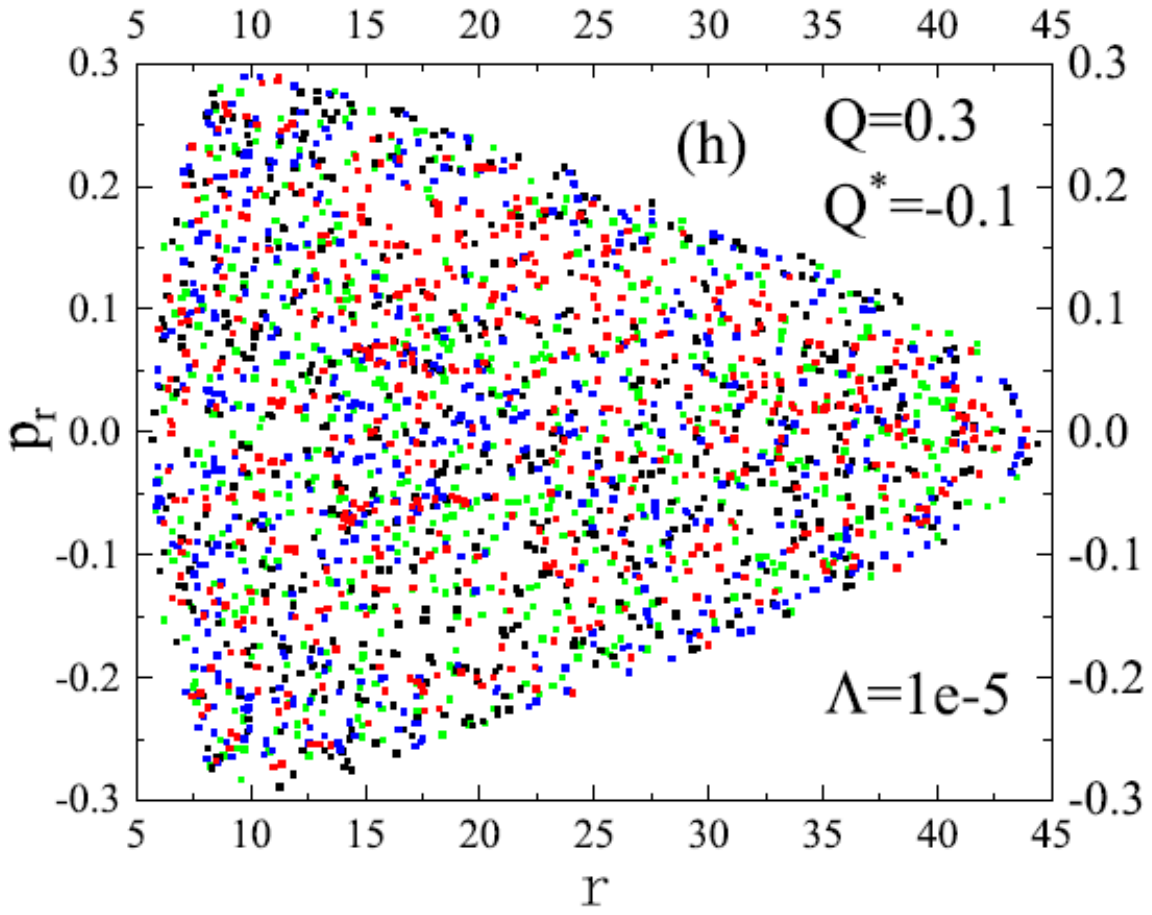}
\includegraphics[scale=0.25]{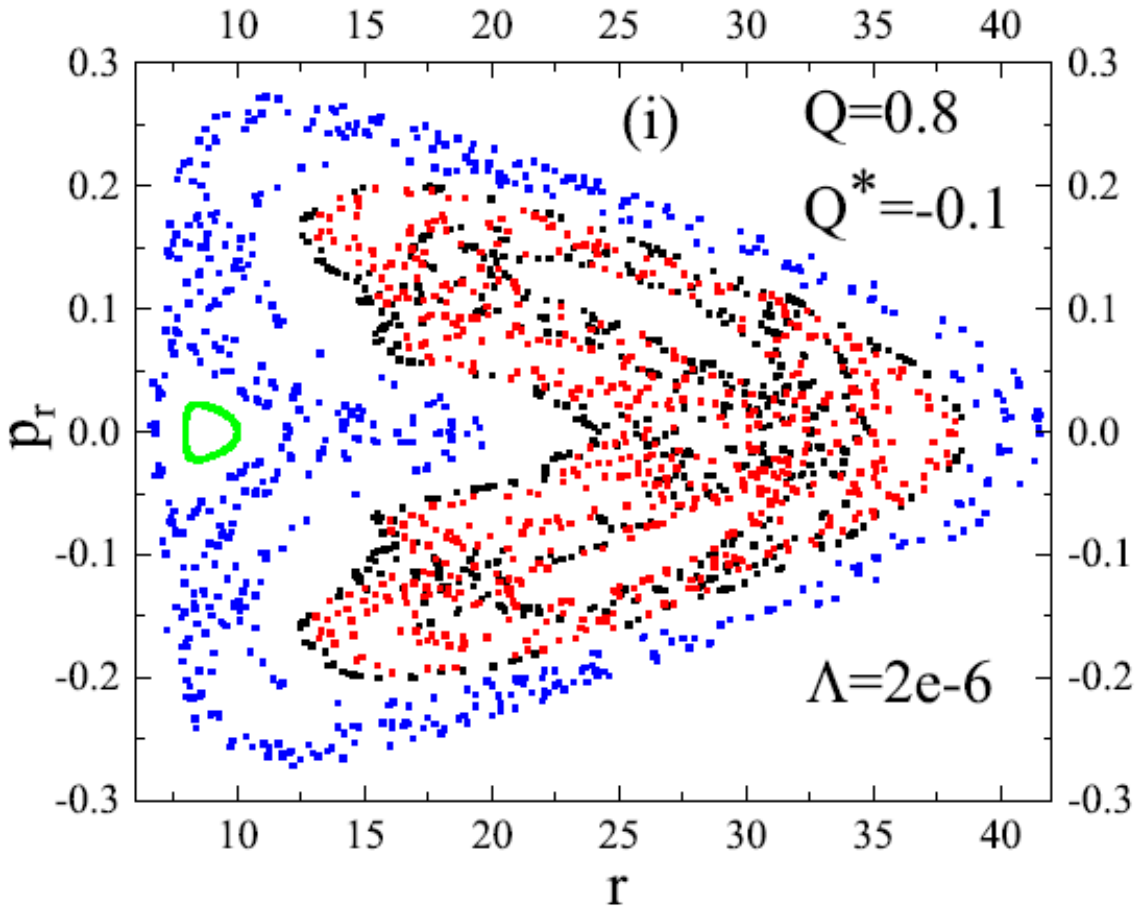}
\includegraphics[scale=0.25]{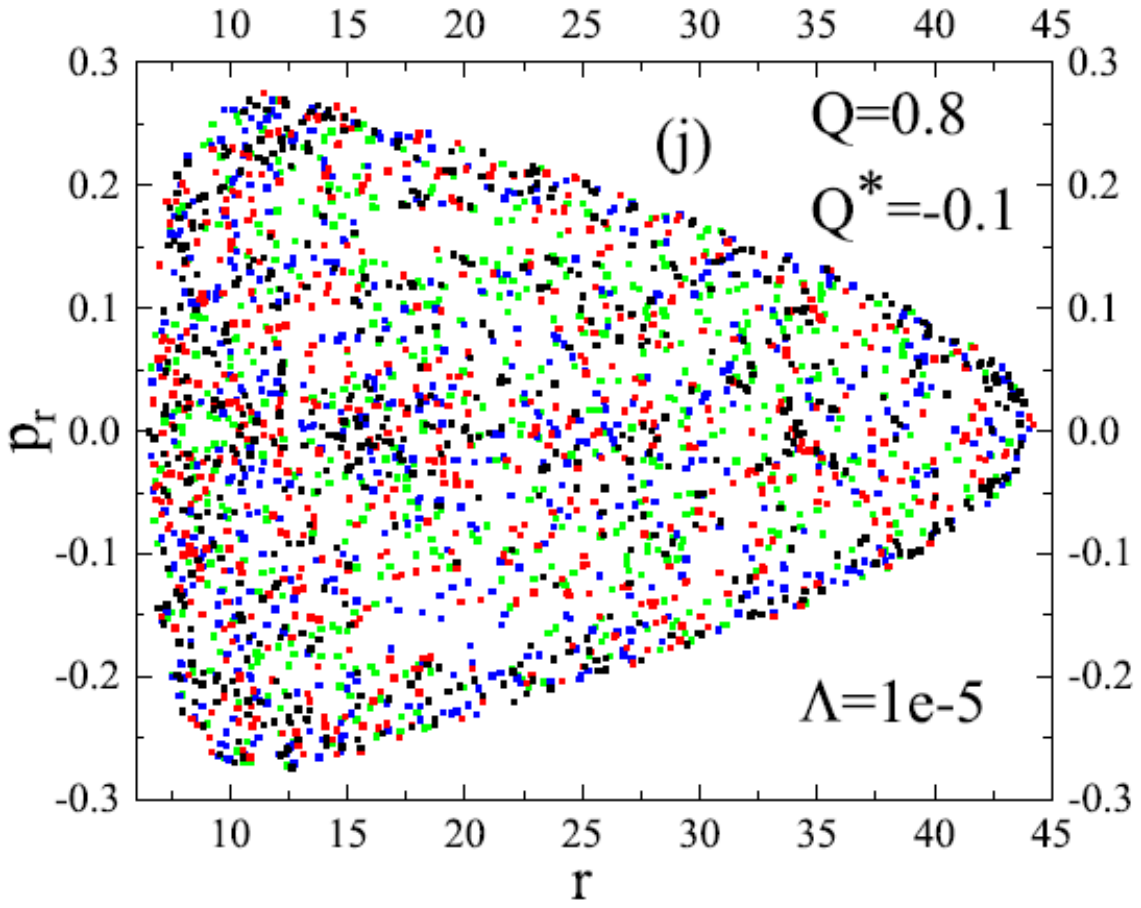}
\includegraphics[scale=0.25]{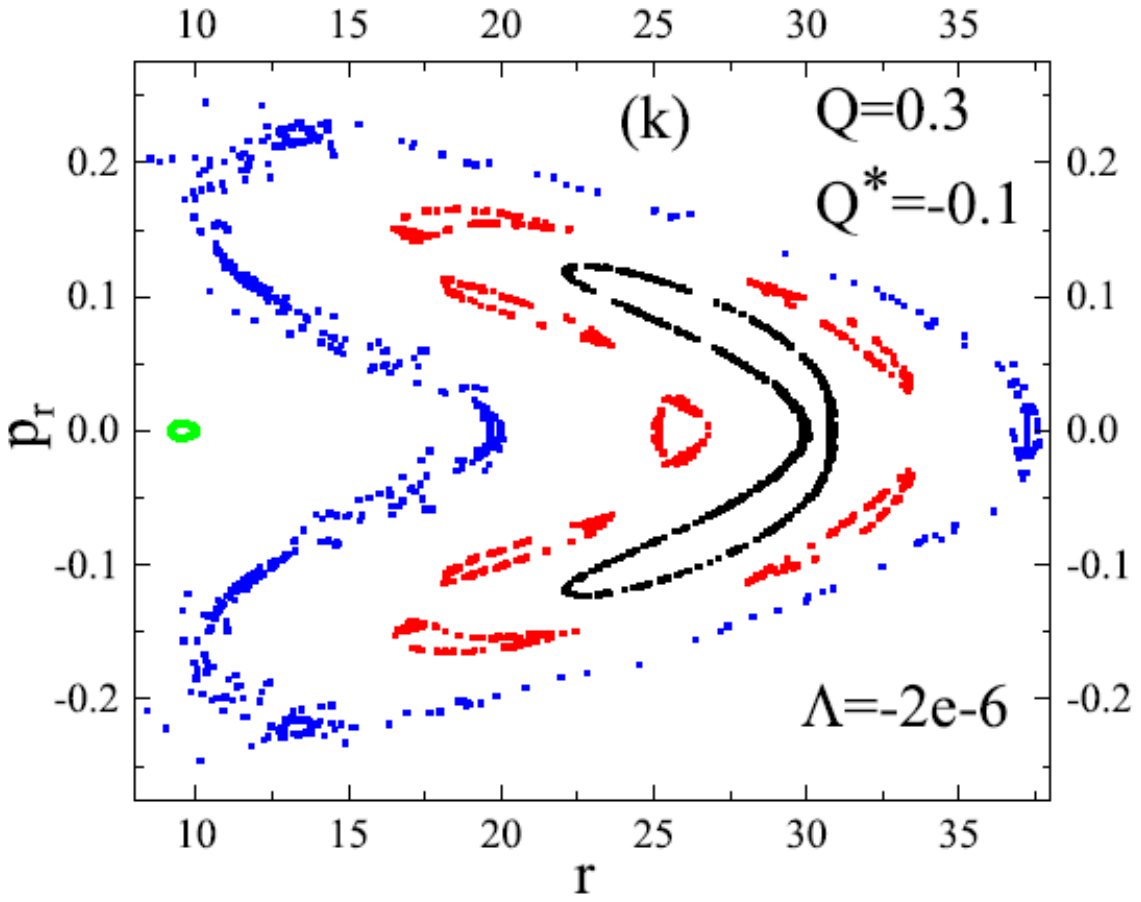}
\includegraphics[scale=0.25]{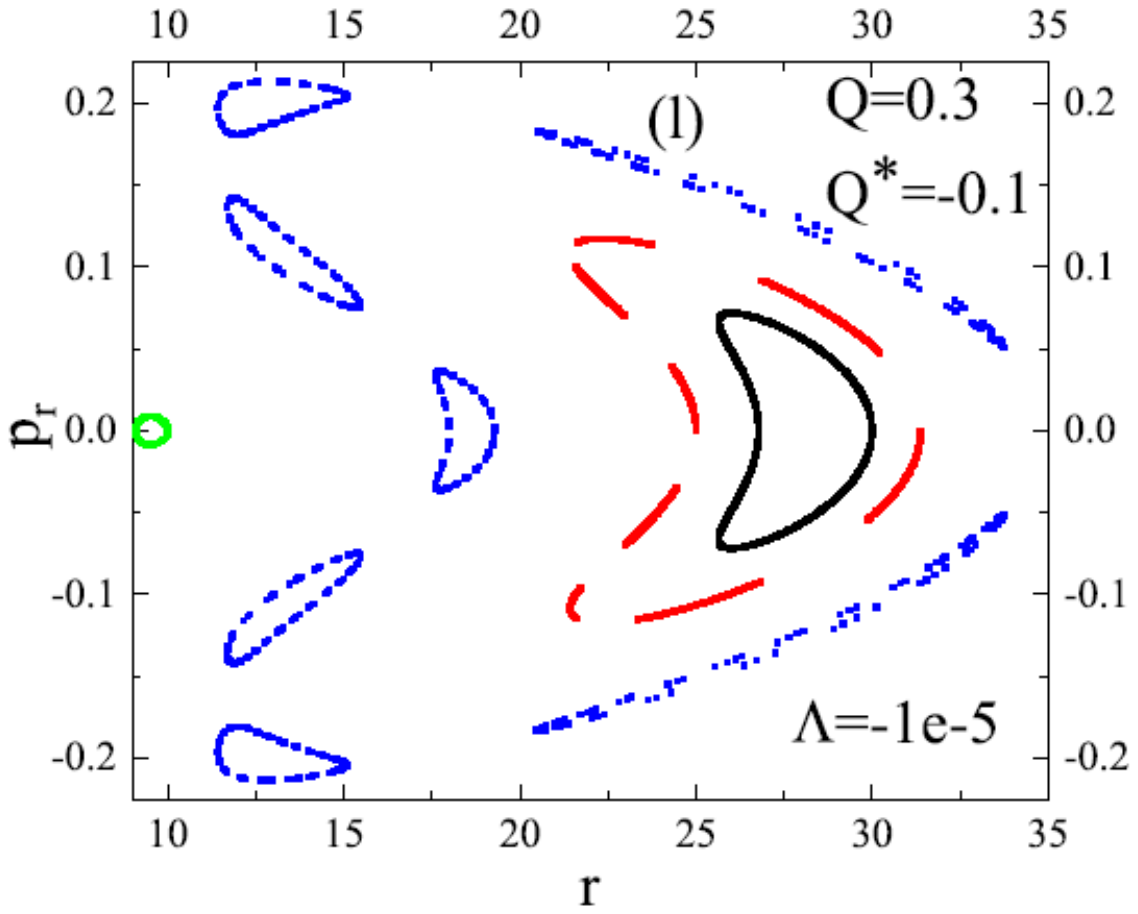}
\caption{Fig. 4 continued.}
 \label{Fig5}}
\end{figure*}


\begin{thebibliography}{}

\bibitem[Abbott et al. (2020a)]{38} Abbott, R.,  Abbott, T. D.,  Abraham, S., et al. 2020a,
ApJL, 900, L13

\bibitem[Abbott et al. (2020b)]{39} Abbott, R.,  Abbott, T. D.,  Abraham, S., et al.  2020b, Phys. Rev.
Lett., 125, 101102

\bibitem[Ade et al. (2016)]{32} Ade, P. A. R.,  Aghanim, N., Arnaud, M., et al. 2016, A$\&$A, 594, A13

\bibitem[Ahmed et al. (2016)]{40} Ahmed,  A. K., Camci, U., $\&$  Jamil, M. 2016, Class. Quant. Grav., 33, 215012

\bibitem[Ashtekar (2017)]{5} Ashtekar, A. 2017, Rep. Prog. Phys., 80, 102901

\bibitem[Ashtekar et al. (2016)]{5b} Ashtekar, A., Bonga, B., $\&$ Kesavan, A. 2016, Phys. Rev. Lett., 116,
051101

\bibitem[Bacchini et al. (2018a)]{11} Bacchini, F., Ripperda, B., Chen, A. Y., $\&$ Sironi, L. 2018a, ApJS, 237, 6

\bibitem[Bacchini et al. (2018b)]{12} Bacchini, F., Ripperda, B., Chen, A. Y., $\&$ Sironi, L. 2018b, ApJS, 240, 40

\bibitem[Brink et al. (2015a)]{12b} Brink, J., Geyer, M., $\&$
Hinderer, T. 2015a,  Phys. Rev. D 91, 083001

\bibitem[Brink et al. (2015b)]{12c} Brink, J., Geyer, M., $\&$
Hinderer, T. 2015b,  Phys. Rev. Lett. 114, 081102

\bibitem[Brown(2006)]{22} Brown, J. D. 2006, Phys. Rev. D, 73, 024001

\bibitem[Carter(1968)]{43} Carter, B. 1968, Phy. Rev., 174, 1559

\bibitem[Deng et al. (2020)]{21} Deng, C., Wu, X., $\&$ Liang, E. 2020, MNRAS, 496, 2946

\bibitem[Dubeibe et al. (2007)]{21a} Dubeibe, F. L., Pach\'{o}n,  L. A., $\&$
Sanabria-G\'{o}mez, J. D. 2007, Phys. Rev. D, 75, 023008

\bibitem[Eatough et al. (2013)]{21b} Eatough, R. P., Falcke, H., Karuppusamy, R., et al. 2013, Nature
, 501, 391

\bibitem[EHT et al. (2019a)]{35} EHT Collaboration, et al. 2019a, ApJL, 875, L1 (Paper I)

\bibitem[EHT et al. (2019b)]{36} EHT Collaboration et al. 2019b, ApJL, 875, L4 (Paper IV)

\bibitem[EHT et al. (2019c)]{37} EHT Collaboration et al. 2019c, ApJL, 875, L6 (Paper VI)

\bibitem[Ernst (1976)]{37a} Ernst, F. J. 1976, J. Math. Phys. 17, 54

\bibitem[Ficek (2015)]{37b} Ficek, F. 2015, Class. Quantum Grav., 32,
235008

\bibitem[Felice \& Sorge (2003)]{44} Felice, D. d, $\&$ Sorge, F. 2003, Class. Quantum Grav., 20, 469

\bibitem[Forest \& Ruth (1990)]{9} Forest, E., $\&$ Ruth, R. D. 1990, Physica D, 43, 105

\bibitem[Fukushima(2003a)]{14} Fukushima, T. 2003a, AJ, 126, 1097

\bibitem[Fukushima(2003b)]{15} Fukushima, T. 2003b, AJ, 126, 2567

\bibitem[Fukushima(2003c)]{16} Fukushima, T. 2003c, AJ, 126, 3138

\bibitem[Fukushima(2004)]{17} Fukushima, T. 2004, AJ, 127, 3638

\bibitem[Gurzadyan(2019)]{33a} Gurzadyan,  V. G. 2019, Eur. Phys. J. Plus, 134,
14

\bibitem[Gurzadyan \& Stepanian(2019)]{33} Gurzadyan,  V. G., $\&$ Stepanian, A. 2019, Eur. Phys. J. C, 79, 169

\bibitem[Han(2008)]{33c} Han, W. B. 2008, Phys. Rev. D, 77, 123007


\bibitem[Hawking \& Page(1983)]{33b}  Hawking, S. W., $\&$ Page, D. N. 1983, Commun. Math. Phys., 87,
577

\bibitem[Hu et al. (2019)]{13} Hu S., Wu X., Huang G., $\&$ Liang E. 2019,
ApJ, 887, 191

\bibitem[Hu et al. (2021)]{13b}  Hu S., Wu X., $\&$ Liang E. 2021,
ApJS, accepted

\bibitem[Jamil(2009)]{13c}  Jamil, M. 2009, Eur. Phys. J. C, 62,
609

\bibitem[Jamil et al. (2008)]{13d}  Jamil, M, Rashid, M. A., $\&$ Qadir, A. 2008, Eur. Phys. J. C, 58,
325

\bibitem[Karkowski1 \& Malec(2014)]{13e}  Karkowski1, J.,  $\&$ Malec, E.
2013, Phys. Rev. D, 87, 044007

\bibitem[Karas \& Vokroulflick\'{y}(1992)]{13f}  Karas, V., $\&$ Vokroulflick\'{y}, D.
1992, Gen. Relativ. Gravit. 24, 729

\bibitem[Kolo\v{s} et al.(2017)]{24a}  Kolo\v{s}, M., Tursunov, A., $\&$ Stuchl\'{i}k, Z. 2017, EPJC, 77, 860


\bibitem[Kop\'{a}\v{c}ek et al.(2010)]{24} Kop\'{a}\v{c}ek, O., Karas, V.,  Kov\'{a}\v{r}, J., $\&$
Stuchl\'{i}k, Z. 2010, ApJ, 722, 1240

\bibitem[Kop\'{a}\v{c}ek \& Karas (2014)]{24b} Kop\'{a}\v{c}ek, O., $\&$ Karas, V. 2014, ApJ, 787,
117

\bibitem[Kop\'{a}\v{c}ek \& Karas (2018)]{24c} Kop\'{a}\v{c}ek, O., $\&$ Karas, V. 2018, ApJ, 853, 53


\bibitem[Kottler(1918)]{4} Kottler, F. 1918,  Ann. Phys., 56, 401

\bibitem[Li \& Wu (2019)]{25a} Li, D., $\&$ Wu, X. 2019,  Eur. Phys. J. Plus, 134,
96

\bibitem[Lubich et al.(2010)]{25} Lubich, C., Walther, B., $\&$ Br\"{u}gmann, B. 2010, Phys. Rev.
D, 81, 104025

\bibitem[Lukes-Gerakopoulos et al.(2016)]{25b} Lukes-Gerakopoulos, G., Katsanikas, M., Patsis, P. A., $\&$ Seyrich, J. 2016, Phys. Rev.
D, 94, 024024

\bibitem[Ma et al.(2008)]{18} Ma, D. Z., Wu, X., $\&$ Zhu, J. F. 2008, New Astrom., 13, 216

\bibitem[Maldacena(1998)]{18b} Maldacena, J. 1998, Adv. Theor. Math. Phys., 2,
231

\bibitem[Mei et al.(2013a)]{27} Mei, L., Ju, M., Wu, X., $\&$ Liu, S. 2013a, Mon. Not. R. Astron. Soc., 435, 2246

\bibitem[Mei et al.(2013b)]{28} Mei, L., Wu, X., $\&$ Liu, F. 2013b, Eur. Phys. J. C, 73, 2413

\bibitem[Nakamura \& Ishizuka (1993)]{3a} Nakamura, Y., $\&$ Ishizuka, T. 1993, Astrophys. Space Sci. 210,
105

\bibitem[Nordstr\"{o}m(1918)]{3} Nordstr\"{o}m, G. 1918, Proc. Kon. Ned. Akad. Wet., 20, 1238

\bibitem[P\'{a}nis et al.(2019)]{3b} P\'{a}nis, R., Kolo\v{s}, M.,  Stuchl\'{i}k, Z.
2019, Eur. Phys. J. C, 79, 479


\bibitem[Perlmutter(1999)]{23a} Perlmutter, S., et al. 1999, ApJ,
517, 565

\bibitem[Preto \& Saha (2009)]{23} Preto, M., $\&$ Saha, P. 2009, ApJ, 703, 1743

\bibitem[Riess et al.(1998)]{23b} Riess, A. G., et al. 1998, AJ, 116,
1009

\bibitem[Reissner(1916)]{2} Reissner, H. 1916, Ann. Phys., 50, 106

\bibitem[Ruth(1983)]{8} Ruth, R. D, 1983, IEEE Trans. Nucl. Sci. NS 30, 2669

\bibitem[Jonathan \& Lukes-Gerakopoulos (2012)]{8a} Jonathan Seyrich, J., $\&$ Lukes-Gerakopoulos, G.  2012, Phys. Rev. D, 86, 124013

\bibitem[Schwarzschild(1916)]{1} Schwarzschild, K. 1916, Stizber. Deut. Akad. Wiss., Berlin, K1.
Math.-Phys. Tech. s., 189

\bibitem[Shafiq(2020)]{42} Shafiq, S., Hussain, S., Ozair, M., Aslam, A., $\&$
Hussain, T. 2020, Eur. Phys. J. C, 80, 744

\bibitem[Shahzad \& Jawad (2019)]{41} Shahzad, M. U., $\&$  Jawad, A. 2019, Can. J. Phys., 97, 742

\bibitem[Stuchl\'{i}k(2005)]{41b} Stuchl\'{i}k, Z. 2005, Modern Physics Letters
A, 20, 561

\bibitem[Stuchl\'{i}k \& Kolo\v{s} (2016)]{41d} Stuchl\'{i}k, Z., $\&$ Kolo\v{s}, M. 2016, Eur. Phys. J. C, 76,
32

\bibitem[Stuchl\'{i}k et al. (2020)]{41e} Stuchl\'{i}k, Z., Kolo\v{s}, M., Kov\'{a}\v{r}, J., $\&$ Tursunov, A. 2020, Universe, 6,
26

\bibitem[Stuchl\'{i}k et al. (2013)]{41c} Stuchl\'{i}k, Z., Kotrlov\'{a}, A., $\&$ T\"{o}r\"{o}k, G. 2013, A$\&$A, 552,
A10

\bibitem[Swope et al. (1982)]{6}  Swope, W. C.,  Andersen, H. C.,  Berens, P. H., $\&$  Wilson, K. R. 1982, J.
Chem. Phys. 76, 637

\bibitem[Takahashi \& Koyama (2009)]{29a} Takahashi, M., $\&$ Koyama, H. 2009, ApJ, 693,
472

\bibitem[Tsang et al. (2015)]{29} Tsang, D.,  Galley, C. R.,  Stein, L. C., $\&$ Turner, A. 2015, ApJL, 809, L9

\bibitem[Tursunov1 et al. (2018)]{29b} Tursunov1, A., Kolo\v{s}, M., Stuchl\'{i}k, Z., $\&$ Gal'tsov,
D. V. 2018, ApJ, 861, 2

\bibitem[Wang et al.(2018)]{20} Wang, S. C., Huang, G. Q., $\&$ Wu, X. 2018, AJ, 155, 67

\bibitem[Wang \& Jiang (2020)]{31b} Wang, X., $\&$ Jiang, J. 2020, JCAP, 07, 052

\bibitem[Wang et al.(2021a)]{30} Wang Y., Sun W., Liu F., Wu X. 2021a, ApJ,
907, 66 (Paper I)

\bibitem[Wang et al.(2021b)]{31} Wang Y., Sun W., Liu F., Wu X., 2021b, ApJ, 909, 22 (Paper
II)

\bibitem[Wang et al.(2016)]{19} Wang, S. C., Wu, X., $\&$ Liu, F. Y. 2016, MNRAS, 463, 1352

\bibitem[Wisdom(1982)]{7} Wisdom, J. 1982, AJ, 87, 577

\bibitem[Wisdom \& Holman (1982)]{10} Wisdom, J., $\&$ Holman, M. 1991, AJ, 102, 1528

\bibitem[Xu \& Wang (2017)]{34} Xu. Z., $\&$ Wang, J. 2017, Phys. Rev. D, 95, 064015

\bibitem[Yi \& Wu (2020)]{45}  Yi, M., $\&$ Wu, X. 2020, Phys. Scr., Phys. Scr., 95, 085008

\bibitem[Yoshida(1990)]{46} Yoshida, H. 1990, Phys. Lett. A, 150, 262

\bibitem[Zelenka et al.(2019)]{46c} Zelenka, O., Lukes-Gerakopoulos, G., $\&$ Witzany, V.
2019, arXiv: 1903. 00360 [gr-qc]


\bibitem[Zelenka et al.(2020)]{46b} Zelenka, O., Lukes-Gerakopoulos, G., Witzany, V., $\&$ Kop\'{a}\v{c}ek, O.
2020, Phys. Rev. D,  101, 024037

\bibitem[Zhong et al.(2010)]{26} Zhong, S. Y., Wu, X., Liu, S. Q., $\&$  Deng, X. F. 2010, Phys. Rev. D, 82, 124040

\end{thebibliography}
\end{document}